\documentclass{JFM-FLM_Au}

\usepackage{array}
\lefttitle{Mundhada et al.}
\righttitle{Mundhada et al.}

\title{A generalized, quasi-universal subgrid model for macroscopic moving contact line flows}

\author{Vyankatesh Manoj Mundhada\aff{1} \and Manoj Kumar Tripathi\aff{1}}

\affiliation{\aff{1}Department of Chemical Engineering, Indian Institute of Science Education and Research Bhopal, Bhopal, India}

\corresau{Manoj Kumar Tripathi, manojkt@iiserb.ac.in}

\begin{document}
\maketitle
\thispagestyle{empty}
\begin{abstract}
Predictions of moving contact line flows are challenging due to the stress singularity associated with the no-slip boundary condition on solid surfaces and the inherently multiscale nature of the phenomenon. Resolving the microscopic length scale renders the numerical simulations intractable for macroscopic flows due to extremely high computational costs. Affordable and predictive numerical solutions for macroscopic flows require a subgrid model, free of case-specific calibration, that mitigates the grid-dependence observed otherwise. We demonstrate a coupling between Cox's matched asymptotic relation and the empirical contact angle models to obtain a quasi-universal subgrid model. The coupling requires a one-time knowledge of the observation length scale or resolution of the experiments used to formulate the empirical models. The present model eliminates the need for prescribing phenomenological parameters such as the microscopic slip length and wall contact angle, allowing for the predictive simulations of such flows. We demonstrate the reduction in error due to grid-dependence by conducting axisymmetric simulations of droplet spreading for a wide range of Reynolds and Weber numbers for three different contact angle models in the advancing and receding contact line motion scenarios. 

\end{abstract}

\begin{keywords}
Moving contact line; Cox theory; Dynamic contact angle

\end{keywords}

\section{Introduction}
\label{sec:intro}
A contact line is a convenient mathematical model for the junction of a solid phase and two immiscible fluid phases in the continuum limit. Contact lines manifest themselves in many multiphase flow scenarios, such as the motion of crude oil and water through rock formations \citep{patel2022computing}, air flow over the mucus lining in the pulmonary airways \citep{gu2013discovery}, and capillary suspensions \citep{bossler2016structure}. The motion of a contact line on a solid surface directly defies the widely assumed and exploited no-slip boundary condition in fluid mechanics. In their landmark paper, \citet{huh1971hydrodynamic} showed the stress singularity that arises due to a no-slip boundary condition, and the necessity of a different kind of boundary condition or a different treatment of the fluid properties at the contact line. Many solutions to relieve the stress singularity at the contact line have since been proposed, including the Navier-slip model \citep{spelt2005level,afkhami2009mesh,ZHAO2020109582,CHAI2021110607}, in which the tangential slip velocity is calculated from the viscous shear stress using a characteristic slip length. Molecular-dynamics simulations by \citet{PhysRevE.68.016306} show that wall slip near the contact line is governed by both viscous stress and uncompensated Young stress. This led to the Generalized Navier Boundary Condition (GNBC), in which the uncompensated Young stress is included as an additional force in the Navier-slip condition. In a sharp-interface description, this force is concentrated at the contact line and is represented using a delta distribution \citep{YAMAMOTO201322, yamamoto2014modeling}. \citet{YAMAMOTO201322} also proposed a simplified form of the GNBC by neglecting the viscous-stress contribution near the contact line. The slip or cutoff lengths associated with these models lie in nanometer scales \citep{SUI201337,yamamoto2014modeling}.

\citet{SHIKHMURZAEV2006121} showed that slip models can retain a pressure singularity at the contact line. In his interface-formation model \citep{shikhmurzaev1994mathematical}, the formation and relaxation of interfaces are explicitly accounted for, yielding a singularity-free description of the experimentally observed rolling kinematics, although at the expense of an increased number of model parameters and greater associated challenges in computational  implementation and cost. Despite these issues, numerical simulations of flows such as sloshing \citep{ming2010numerical}, coating \citep{zhan2023numerical}, and other multiphase flows \citep{sharma2021regimes}, without any special treatment of the contact angle boundary conditions, are routine. In several of these studies, the errors due to the inaccurate modeling are apparently low enough, as evident from the experimental validations presented therein, for the flow regimes and parameters considered. This is one of the topics discussed in the present work.

Nevertheless, the problems associated with the numerical modeling of flows with moving contact lines have been recognized, and there has been an effort to improve the subgrid models for the phenomenon, mostly in the past two decades \citep{afkhami2009mesh,yokoi2009numerical,legendre2015comparison}. Majority of the methods employ a contact angle model to calculate the contact angle to be imposed at the contact line. These contact angle models relate the contact line velocity with the angle, empirically. Thus an accurate computation of the contact-line velocity has received significant attention in the literature. The most common method adopted in the literature is to assign the velocity of the cell containing the contact line as the contact line velocity \citep{spelt2005level,malgarinos2014vof,gohl2018immersed}. However, due to the specific flow profile at the three-phase junction, this assumption gives rise to errors whose magnitude is highly sensitive to the contact angle \citep{roisman2008drop}. Interpolation methods to estimate the contact line velocity from neighboring cell-center velocities have improved the accuracy of such methods against the experimental observations \citep{roisman2008drop,han2025geometric}. Accurate estimation of the contact line velocity solves only one of the problems associated with the modelling of the subgrid physics of the moving contact lines.

One of the seemingly insurmountable challenges has been to make such simulations grid-independent\citep{afkhami2009mesh,legendre2015comparison}. The root of this obstacle lies in the physics at the nano- and microscale at the contact line. Grid convergence in moving contact line simulations is readily achieved when the computational mesh is fine enough to resolve the nanoscale slip length in conjunction with a slip model \citep{fullana2024consistent}. However, such a simulation methodology is prohibitively expensive for macroscopic flow applications, where the computational domain is several millimeters or centimeters in size. Several methods in the level-set \citep{spelt2005level, SUI201337}, volume of fluid (VOF) \citep{afkhami2008,vsikalo2005dynamic, DUPONT20102453,renardy2001numerical}, diffuse-interface \citep{ding2007wetting,LIU2015484}, and front-tracking \citep{ZENG2025114259,JANSSEN2024113449,mirsandi2018numerical,shang2018gnbc,shang2019numerical} frameworks have been proposed for macroscopic simulations of these flows; most of which discuss the accurate imposition of a known contact angle as a boundary condition, with only a few that discuss the issue of grid-independence. The relationship between the macroscopic (apparent) and microscopic (inner) length scales and contact angles, and the contact line speed due to \citet{cox1986dynamics} has been invariably employed in the literature pertaining to the issue of grid-convergence in such simulations \citep{SUI201337,afkhami2009mesh}. Cox's relation has notably been leveraged to arrive at a subgrid model which shows scale-invariance and implemented in a VOF framework by \citet{afkhami2009mesh}. In this approach, the results at any grid size can be made to match the solution at a reference grid size by choosing a suitable relationship between the numerically imposed contact angle and the grid size. Moreover, to achieve a match between the simulations and experiments, the method requires a fitting parameter in the model that is obtained in conjunction with the experimental data.

Cox's asymptotic matching is key to obtaining a subgrid model for contact angle, because it shows that the angle made by the position vector to any point on the interface depends on the vector's length, also known as viscous bending of the interface \citep{voinov1976hydrodynamics}. However, Cox's asymptotic matching between the three length scales, namely, the inner, the intermediate, and the outer scales, introduces phenomenological parameters such as the slip length and the microscopic contact angle that need to be known in order to calculate the contact angle at any distance away from the wall. As noted in the literature \citep{afkhami2018}, the predictive ability of simulations and the applicability of such a scale-invariant model are limited due to the lack of knowledge of these microscopic parameters. Formulations of boundary conditions for macroscopic contact line flows that do not involve specifying a contact angle or slip length have been developed in the literature, such as the curvature boundary condition \citep{luo2016curvature} and the contact line force method \citep{vsikalo2005dynamic,esteban2023contact}. Although different in their origin, both of these categories of methods assume the computational grid size to be of the order of the outer region's length scale. Moreover, when the empirical contact angle models are employed in conjunction with these methods, the models are assumed to be applicable at the grid length scale. Some of these studies show grid-independence of the equilibrium shape of spreading droplets rather than for the transient spreading phase \citep{esteban2023contact}, while the others show only experimental validations in place of a grid-independence test. While this elucidates the ability of the numerical method to simulate the shown cases, the fundamental problem of grid-dependence in such simulations remains.  

\begin{figure}
    \centering
      \includegraphics[width=10cm]{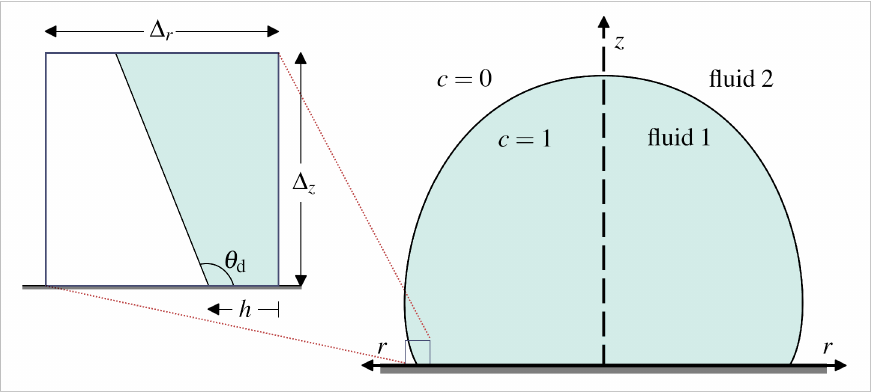}
    \label{fig:wide}
    \caption{An axisymmetric liquid (fluid `1') droplet on a solid substrate in the presence of fluid `2' in $(r,z)$ coordinates. The inset shows a region of size $\Delta_r\times\Delta_z$ containing the contact-line, which may represent a discrete finite volume, `contact-line cell'.}
\label{fig:innerouter}
\end{figure}

In this work, we leverage the observed reality of the empirical contact angle models and the microscopic-to-macroscopic bridge provided by the Cox theory to derive a scale-invariant subgrid model that may facilitate grid-independent simulations of moving contact line flows. We show that our model does not require a prescription of phenomenological parameters such as the microscopic slip length or wall contact angle. Moreover, the proposed method naturally allows for predictive simulations of such flows as it does not require any fitting parameters. Without the loss of generality, we consider an axisymmetric droplet as shown in Fig.~\ref{fig:innerouter}, as the prototype system to represent the flows with moving contact lines. The rest of the paper is organized as follows. We present the governing equations and the numerical method to simulate moving contact line dynamics of a droplet on a flat plate in Sec. \ref{sec:num}, followed by an investigation of grid-dependence in the light of the Cox theory in Sec. \ref{sec:thetad}. Subsequently, we develop `modified' contact angle models as subgrid models for the dynamic contact angle in Sec. \ref{sec:coupling_cox}, show the grid-independent results from the implementation of these models for a wide range of $Re$, $We$, and contact angles in Sec. \ref{Results_discussion} and conclude the findings in Sec. \ref{sec:conclusion}.     

\section{Governing Equations and Numerical Framework}
\label{sec:num}
\subsection{Governing Equations}
We consider a system composed of two immiscible, incompressible, and Newtonian
fluids whose densities are $\rho_1$ and $\rho_2$, and dynamic viscosities $\mu_1$ and
$\mu_2$. We solve a single set of mass and momentum conservation equations over
the entire domain, with density and viscosity treated as spatially and temporally varying
fields. The equations are solved in a cylindrical ($r$,$z$) coordinate system with an axisymmetric constraint about the $z$-axis. The governing equations, non-dimensionalized with the characteristic length
$L$, velocity $U_0$, density $\rho_1$, viscosity $\mu_1$, time $L/U_0$, and pressure $\rho_1U_0^2$, are as follows.
\begin{eqnarray}
\nabla \cdot \mathbf{u} &=& 0 ,
\label{eq:cont}
\\[6pt]
\frac{\partial (\rho \mathbf{u})}{\partial t}
+ \nabla \cdot (\rho \mathbf{u} \mathbf{u})
&=& -\nabla p
+ \frac{1}{\mathrm{Re}} \nabla \cdot \boldsymbol{\tau}
+ \frac{1}{\mathrm{Fr}^2} \rho \mathbf{g}
+ \frac{1}{\mathrm{We}} \mathbf{f}_\sigma ,
\label{eq:mom}
\end{eqnarray}
where $\mathbf{u}$, $p$, $\rho$, $\mu$, $\mathbf{g}$, and $\mathbf{f}_\sigma$ denote the dimensionless velocity vector, pressure, density, viscosity, gravitational acceleration, and the interfacial-tension force per unit volume acting at the fluid-fluid interface expressed as a body force \citep{brackbill1992continuum,saifi2022thermocapillary}. The dimensionless viscous stress tensor is
\begin{equation}
\boldsymbol{\tau}
=
\mu
\left(
\nabla \mathbf{u}
+ (\nabla \mathbf{u})^\top
\right).
\end{equation}
The Reynolds, Weber, and Froude numbers are defined as
\begin{equation}
\mathrm{Re} = \frac{\rho_1 U_0 L}{\mu_1}, \qquad
\mathrm{We} = \frac{\rho_1 U_0^2 L}{\sigma}, \qquad
\mathrm{Fr} = \frac{U_0}{\sqrt{gL}},
\end{equation}
where $\sigma$ is the interfacial-tension coefficient.
A scalar field $c$ is defined, which denotes the volume fraction of the outer fluid (fluid $1$), varying abruptly from $0$ (fluid $2$) to $1$ (fluid $1$). The advection equation for the volume fraction field then becomes: 
\begin{equation}
    \frac{\partial c}{\partial t} + \textbf{u} \cdot \nabla c = 0.
    \label{eq:vofadv}
\end{equation}
A numerical solution of the abovementioned equations in the axisymmetric cylindrical coordinate system is discussed next.

\subsection{Numerical Framework}

Eqs.~\eqref{eq:cont}, \eqref{eq:mom} and \eqref{eq:vofadv} are solved numerically using a finite-volume method with a volume-of-fluid (VoF) framework, implemented in the open-source \textit{Basilisk} code (\url{http://basilisk.fr/}) \citep{popinet2009}. The quadtree-based adaptive mesh refinement approach in this code has been leveraged to provide a higher spatial resolution at the fluid-fluid interface and near the contact line in our simulations. 

Instead of a sudden jump in $c$ and the fluid properties, we introduce their smooth variation by assuming fractional values of $c$ in the computational cells containing the interface. The dimensionless density $\rho$ and viscosity $\mu$ fields are assumed to be linear functions of $c$:
\begin{align}
\rho &= c + (1-c)\rho_r, \\
\mu &= c + (1-c)\mu_r,
\end{align}
where $\rho_r$ and $\mu_r$ are the density and viscosity ratios, defined as $\rho_2/\rho_1$ and $\mu_2/\mu_1$, respectively. A fractional-step projection method is used to numerically solve the governing equations. In this method, an intermediate velocity field is first computed by solving the momentum equation, wherein the velocity advection term is evaluated using a second-order unsplit upwind Bell–Colella–Glaz scheme \citep{bell1989second}. The divergence of this intermediate velocity field is then incorporated as the source term in a Poisson equation for the pressure field. Pressure correction is applied to project the velocity field onto a divergence-free space. Averaging these cell-centered velocities yields face-centered velocities, which are then used to advect volume fractions. Surface tension effects are incorporated by the scheme proposed by \citet{popinet2009}, which combines a geometrical VOF method with a balanced-force continuum surface force formulation. The interface normal and curvature are computed using a generalized height-function method, specifically adapted for quadtree grids, employing asymmetric stencils. When heights are missing, interface normals are calculated using the mixed-Youngs-centered scheme, ensuring high accuracy near non-uniform grids and solid boundaries. A number of test cases available with the solver's source code, and validation cases specific to gas-liquid and liquid-liquid flows presented in the literature \citep{Sakakeeny2020,sakakeeny2021a,kulkarni2023,chen2025} justify the choice of the numerical method and the specific code, \textit{Basilisk}.

The present work focuses on macroscopic flow configurations involving moving contact lines, for which the computational cell size is orders of magnitude much larger than the slip length. Under these conditions, imposing a no-slip boundary condition along the entire solid surface, including at the contact line, is physically justified. Moreover, it has been shown that to determine the macroscopic dynamics, it is sufficient to apply the observed contact angle boundary condition beyond a cutoff region, without mandating the exact microscopic mechanism by which the contact line moves \citep{kistler1993hydrodynamics,afkhami2018}. Thus, another important boundary condition in such problems is the dynamic contact angle, $\theta_d$. For a known value of $\theta_d$ at the cell-center, the boundary conditions for the tangential and normal components of the height functions are modified near the contact line following \citet{afkhami2008}, such that the resulting interface reconstruction yields the prescribed contact angle. 

It should be noted that the numerically imposed $\theta_d$ is different from the microscopic wall contact angle ($\theta_w$) as we do not resolve the length-scales smaller than the grid-size. The dynamic contact angle generally varies in time and must therefore be updated at every time step. In the following, we discuss the evaluation of $\theta_d$.

\subsubsection{Calculation of $\theta_d$}
\label{subsubsec:theta_d}
To begin with, at each time-step, we determine whether the contact line is pinned by evaluating the hysteresis contact angle, $\theta_{d,\mathrm{hys}}$, using the approach described by \citet{FangEtAl2008} and previously implemented in our previous work \citep{saifi2022thermocapillary}. If $\theta_{d,\mathrm{hys}}$ lies between the static receding and advancing contact angles, $\theta_r$ and $\theta_a$, respectively, the contact line is treated as pinned, and the dynamic contact angle is taken as $\theta_d=\theta_{d,\mathrm{hys}}$. For $\theta_{d,\mathrm{hys}} > \theta_a$, the contact line is classified as advancing, while for $\theta_{d,\mathrm{hys}} < \theta_r$, it is classified as receding. In these cases, the dynamic contact angle is evaluated using the corresponding theoretical or empirical contact angle model. The dynamic contact-angle models considered in the present study are summarized in Table~\ref{tab:models}. The general form of contact angle models can be written as follows.

\begin{equation}
\theta_{d} = \Theta(\mathrm{Ca}_{\mathrm{CL}}),
\label{eq:dynamic_contact angle}
\end{equation}
which relates the contact angle to the contact-line capillary number
$\mathrm{Ca}_{\mathrm{CL}}$, defined as
\begin{equation}
\mathrm{Ca}_{\mathrm{CL}} = \frac{\mu_1U_0}{\sigma} u_{\mathrm{CL}},
\end{equation}
where $u_{\mathrm{CL}}$ represents the dimensionless contact-line velocity.

\begin{table}
  \begin{center}
\def~{\hphantom{0}}
\small
\renewcommand{\arraystretch}{1.15}
\setlength{\tabcolsep}{4pt}

\begin{tabular}{p{2.6cm} @{\hspace{3pt}} p{2.0cm} @{\hspace{2pt}} p{9.6cm}}
    \textrm{Model} &
      \textrm{Scenario} &
      \textrm{Dynamic contact-angle relation} \\[3pt]

      \citet{kistler1993hydrodynamics} &
      Advancing \par
      and \par 
      receding &
      $\displaystyle
      \theta_d=f_H\!\left[Ca+f_H^{-1}(\theta_e)\right].
      $
      \par\smallskip
      where $f_H(x)$ is Hoffman's empirical function~\citep{hoffman1975study}, given by
      \par\smallskip
      {\centering
      $\displaystyle
      f_H(x)=\cos^{-1}\!\left\{
      1-2\tanh\!\left[
      5.16\left(
      \frac{x}{1+1.31x^{0.99}}
      \right)^{0.706}
      \right]
      \right\},
      $
      \par}
      \smallskip
      with $\theta_e=\theta_a$ and $\theta_r$ for an advancing and a receding contact line, respectively.
      \\[4pt]

      \citet{jiang1979correlation} &
      Advancing &
      $\displaystyle
      \theta_d=
      \cos^{-1}\!\left[
      \cos\theta_a-
      (1+\cos\theta_a)\tanh(4.96\,Ca^{0.702})
      \right]
      $
      \\[2pt]

      \citet{tanner1979spreading} &
      Receding &
      $\displaystyle
      \theta_d^{\,3}
      =
      \theta_r^{\,3}
      +72\,Ca
      $
      \\

  \end{tabular}

  \caption{Dynamic contact-angle models employed in the present study.}
  \label{tab:models}
  \end{center}
\end{table}
For simplicity, we consider an axisymmetric configuration in cylindrical coordinates, in which the contact line is either a point or a set of discrete points. Without loss of generality, only a single contact point is considered in the following discussion. In this case, the contact-line velocity is obtained directly from the contact-line position, $r_{\mathrm{CL}}$, by differentiating it with respect to time, i.e., $u_{\mathrm{CL}} = dr_{\mathrm{CL}}/dt$. As mentioned in the introduction (Sec. \ref{sec:intro}), several methods for calculating the contact line velocity have been proposed in the literature, calculating it from the contact line position is the most direct and convenient method when the position is represented by a point. Thus, the contact-line position, $r_{\mathrm{CL}}$, becomes an additional unknown that needs to be determined. The analytical relationship between the volume fraction $c$ and the geometrical parameters of the reconstructed interface, which includes $r_{\mathrm{CL}}$, in the Cartesian coordinates has been derived by \citet{scardovelli2000}, for which a correction in the cylindrical coordinates was subsequently given by \citet{burevs2021piecewise}.

Consider a computational cell of dimensions $\Delta_r$ and $\Delta_z$ with volume fraction $c$ and the equation of the linear interface in the contact-line cell to be $z'\tan\theta_d + r' = h$,
where $r'$ and $z'$ denote the local coordinates measured from the inner corner of the contact-line cell on the wall (as shown in the inset of Fig.~\ref{fig:innerouter}). The corresponding volume fraction in the computational cell is given by

\begin{equation}\label{Fun1}
\begin{split}
c = \left(\frac{r_c}{r_{cc}}\right)\frac{\tan\theta_d}{2\Delta_r\Delta_z}
\Big[
h^2
- H(h-\Delta_r)(h-\Delta_r)^2
- H\!\left(h-\frac{\Delta_z}{\tan\theta_d}\right)
\left(h-\frac{\Delta_z}{\tan\theta_d}\right)^2
\Big],
\end{split}
\end{equation}

where $H(x)$ denotes the Heaviside function,

\begin{equation}
H(x)=
\begin{cases}
0, & x<0,\\
1, & x\ge 0.
\end{cases}
\end{equation}

Here, $r_c$ is the radial coordinate of the centroid of the cross-sectional area occupied by fluid 1 in the $r$-$z$ plane, and $r_{cc}$ is the radial coordinate of the cell center. The contact-line position is then obtained as 
\begin{equation}
\label{eq:rcl}
r_{\mathrm{CL}} = r_{cc} - \frac{\Delta_r}{2} + h.
\end{equation}

For $r_{cc} \gg \Delta_r$, $r_c \approx r_{cc}$, and Eq.~(\ref{Fun1}) reduces to the expression of \citet{scardovelli2000} used in Cartesian coordinates. In the present study, the contact line remains sufficiently far from the axis of symmetry such that this approximation is valid.

In the literature, interface reconstruction is typically performed using the contact angle from the previous time step to evaluate the contact-line position, either directly to determine the contact-line speed or indirectly by extrapolating cell-centered velocities to the contact line \citep{roisman2008drop}. Consequently, the evaluation of the contact-line speed generally relies on temporally staggered quantities. In the present work, $r_{\mathrm{CL}}$, $u_{\mathrm{CL}}$, and $\theta_d$ are evaluated simultaneously at the current time step, thereby maintaining consistency within the VOF framework and avoiding errors associated with temporally staggered information. To the best of our knowledge, a simultaneous evaluation of these quantities within the present framework has not been reported previously.

For a given time-step, the local interface intercept $h$ is calculated with respect to a common reference location, such that $dr_{\mathrm{CL}}/dt = dh/dt$. Moreover, the
contact-line velocity is assumed to vary linearly in time over a single time
step. This yields the discretized expression for the capillary number at the next time step as follows.

\begin{equation}
\mathrm{Ca}_{\mathrm{CL}}^{\,n+1}
=
\frac{\mu_1U_0}{\sigma}\,u_{\mathrm{CL}}^{\,n+1}
=
\frac{\mu_1U_0}{\sigma}
\left[
2\left(\frac{h^{\,n+1}-h^{\,n}}{\Delta t}\right) - u_{\mathrm{CL}}^{\,n}\right],
\label{eq:Ca_np1}
\end{equation}

where $\Delta t$, $h^{n}$, $h^{n+1}$, and $u_{\mathrm{CL}}^{\,n}$ are the time step size, the intercept at the current and the next time step, and the contact line velocity at the current time step, respectively. Note that $h^{n}$ and $h^{n+1}$ are measured from the same reference point. This capillary number is used as an input to the dynamic contact-angle model, expressed in the discretized form as follows.

\begin{equation}
\theta_{d}^{\,n+1}
=
\Theta\!\bigl(\mathrm{Ca}_{\mathrm{CL}}^{\,n+1}(h^{\,n+1})\bigr).
\label{eq:theta_of_h}
\end{equation}

From Eqs. ~\eqref{Fun1} and ~\eqref{eq:theta_of_h}, we obtain a single nonlinear equation in $h^{\,n+1}$ as follows.

\begin{equation}
\begin{split}
\label{eq:vf_h_only}
F(h^{n+1})
&=
\frac{\tan\!\bigl(\Theta(h^{\,n+1})\bigr)}{\Delta_r\Delta_z}
\Bigg[
\left(h^{\,n+1}\right)^2
-
H\!\left(h^{\,n+1}-\Delta_r\right)
\left(h^{\,n+1}-\Delta_r\right)^2 \\[2pt]
&\quad -
H\!\left(
h^{\,n+1}
-
\frac{\Delta_z}{\tan\!\bigl(\Theta(h^{\,n+1})\bigr)}
\right)
\left(
h^{\,n+1}
-
\frac{\Delta_z}{\tan\!\bigl(\Theta(h^{\,n+1})\bigr)}
\right)^2
\Bigg] - c^{\,n+1} = 0,
\end{split}
\end{equation}

A Newton--Raphson iteration for this equation, for an iteration number $i+1$, can be written as follows.

\begin{equation}
\label{eq:NR_h}
h_{i+1}^{\,n+1}
=
h_i^{\,n+1}
-
\frac{F(h_i^{\,n+1})}
{\left.\dfrac{dF}{dh}\right|_{h_i^{\,n+1}}},
\end{equation}
where the initial guess is taken as $h_0 = h^{\,n}$. The derivative $\mathrm{d}F/\mathrm{d}h$ in Eq. ~\eqref{eq:NR_h} is obtained by direct differentiation of the residual
$F(h)$, while treating the
Heaviside terms consistently, with the following mathematical expression.

\begin{equation}\label{df_dh}
\begin{array}{rl}
\displaystyle
\frac{dF}{dh} ={}&
\displaystyle
\frac{\sec^2\Theta}{2\Delta_r\Delta_z}
\frac{d\Theta}{dCa_{\mathrm{CL}}}
\frac{dCa_{\mathrm{CL}}}{dh}
\left[
h^2
- H(h-\Delta_r)(h-\Delta_r)^2
- H\!\left(h-\frac{\Delta_z}{\tan\Theta}\right)
\left(h-\frac{\Delta_z}{\tan\Theta}\right)^2
\right]
\\[6pt]
&\displaystyle
+\frac{\tan\Theta}{\Delta_r\Delta_z}
\left[
h-H(h-\Delta_r)(h-\Delta_r)
\right.
\\[6pt]
&\displaystyle\left.
\quad
- H\!\left(h-\frac{\Delta_z}{\tan\Theta}\right)
\left(h-\frac{\Delta_z}{\tan\Theta}\right)
\left(
1+\Delta_z\csc^2\Theta
\frac{d\Theta}{dCa_{\mathrm{CL}}}
\frac{dCa_{\mathrm{CL}}}{dh}
\right)
\right].
\end{array}
\end{equation}

Moreover, we employ the following discretized expression for the derivative $dCa_{\mathrm{CL}}/dh|_{h_i^{\,n+1}}$.

\begin{equation}
\label{eq:dCa_dh_final}
\left.\frac{d\mathrm{Ca_{\mathrm{CL}}}}{dh}\right|_{h_i^{\,n+1}}
=
\frac{2\mu_1U_0}{\sigma\Delta t}\,
\left[\frac{
\left(
h_i^{\,n+1}-h^{\,n}-u_{\mathrm{CL}}^{\,n}\Delta t
\right)}
{
2\left(
h_i^{\,n+1}-h^{\,n}
\right)-u_{\mathrm{CL}}^{\,n}\Delta t
}\right].
\end{equation}

The preceding development is purely kinematic and does not depend on the
specific functional form of the dynamic contact-angle relation
$\Theta(Ca_{\mathrm{CL}})$. Consequently, the formulation applies to any dynamic
contact-angle model, provided that the derivative
$d\Theta/dCa_{\mathrm{CL}}$ can be evaluated (analytically or numerically). Once the
contact-line position $h^{\,n+1}$ has been determined, the corresponding contact
angle is recovered directly from Eq.~\eqref{eq:theta_of_h}. Having established the framework for incorporating the contact-line boundary condition into the numerical method, we now proceed to verify and validate the proposed formulation. 

\subsection{Validation and verification of the numerical method}
\label{sec:verification}
A numerical method needs to be verified for grid-sensitivity and validated against known results. The present numerical method has already been verified for two-phase flows without contact lines as mentioned in Sec. \ref{sec:num}. Therefore, we perform a validation against a known experimental result (case A) and then attempt to study the grid dependence of the method for this case and another set of parameters (case B).

To quantify grid-independence, we define the percentage difference in the interface shape, $E_s$ by using the mean nearest-interface distance, similar to the chamfer distance defined for fluid-interface comparisons by \citet{HASHEMI2024116699}. Each interface is represented by a set of $N=1000$ points sampled uniformly on it, named $\mathcal{P}_{\mathrm{ref}} (= \{\boldsymbol{x}_i^{ref}\})$ for $i=1:N$ and $\mathcal{P} (= \{\boldsymbol{x}_i\})$ for $i=1:N$ for the reference and the current grid, respectively, and the shortest Euclidean distance from each point to the other interface is calculated. The percentage shape difference is defined as
\begin{equation}
E_s(t )
=
\frac{1}{2N}
\left[
\sum_{i=1}^{N}
\left\|
\boldsymbol{x}_i-\mathcal{P}_{\mathrm{ref}}
\right\|
+
\sum_{j=1}^{N}
\left\|
\boldsymbol{x}^{\mathrm{ref}}_j-\mathcal{P}
\right\|
\right]
\times 100,
\label{eq:shape_error}
\end{equation}
where $\|\boldsymbol{x}-\mathcal{P}\|$ is the shortest Euclidean distance from $\boldsymbol{x}$ to $\mathcal{P}$.Here, the distances are evaluated in both directions and averaged to obtain a symmetric measure.

To quantify the pointwise relative percent error in the contact-line position, we define $E_r$ as
\begin{equation}
E_r(t)
=
\frac{\left|r_{CL}(t)-r_{CL,\mathrm{ref}}(t)\right|}
     {r_{CL,\mathrm{ref}}(t)}
\times 100.
\end{equation}
where $r_{CL,\mathrm{ref}}(t)$ is the contact-line position for the reference grid. 

For a quantity $\Psi$, the mean ($\Psi^{L_1}$) and maximum ($\Psi^{L_{\infty}}$) in an interval of interest are then defined as $\frac{1}{N} \sum_{n=1}^{N} \Psi(t_n)$ and $\max_{1\le n\le N}\Psi(t_n)$, respectively, where $\{t_n\}_{n=1}^{N}$ denotes a discrete set of uniformly sampled time instants over the interval.

\subsubsection{Cases A: Gas-liquid system}
\label{subsec:gasliquid}
To validate the methodology, simulations are performed for droplet impact under gravity, corresponding to case 1 of the experiments of~\citet{roisman2008drop}. The boundary conditions used in the present simulations match those of the numerical simulations in the reference. Initially, we keep the droplet away from the solid boundary with its closest point at a distance of two computational cells from the boundary and assign the entire droplet a downward velocity such that it gains the impact velocity given in the reference when it touches the solid surface. This is done by estimating the droplet's kinetic energy at the initial location from the known potential energies and the kinetic energy at the impact point. The corresponding non-dimensional parameters for the problem, with impact velocity as the velocity scale, are $\rho_r = 0.001$, $\mu_r = 0.0018$, $Re=400$, $We=0.87$, and $Fr=0.4$, along with the advancing and receding contact angles $\theta_A=125^\circ$ and $\theta_R=65^\circ$, respectively. The advancing angle given in the reference is $120^\circ$; however, upon inspecting the experimental data of the observed contact angle and contact-line position, the advancing angle is found to be $125^\circ$ for the Kistler model to be valid. The simulations are carried out in an axisymmetric domain using adaptive mesh refinement with different maximum grid-sizes, $\Delta_i=D_0/2^i$ where $i\in\{5,6,7,8\}$. 

\begin{figure}[htpb]
\centering
\includegraphics[width=0.6\columnwidth]{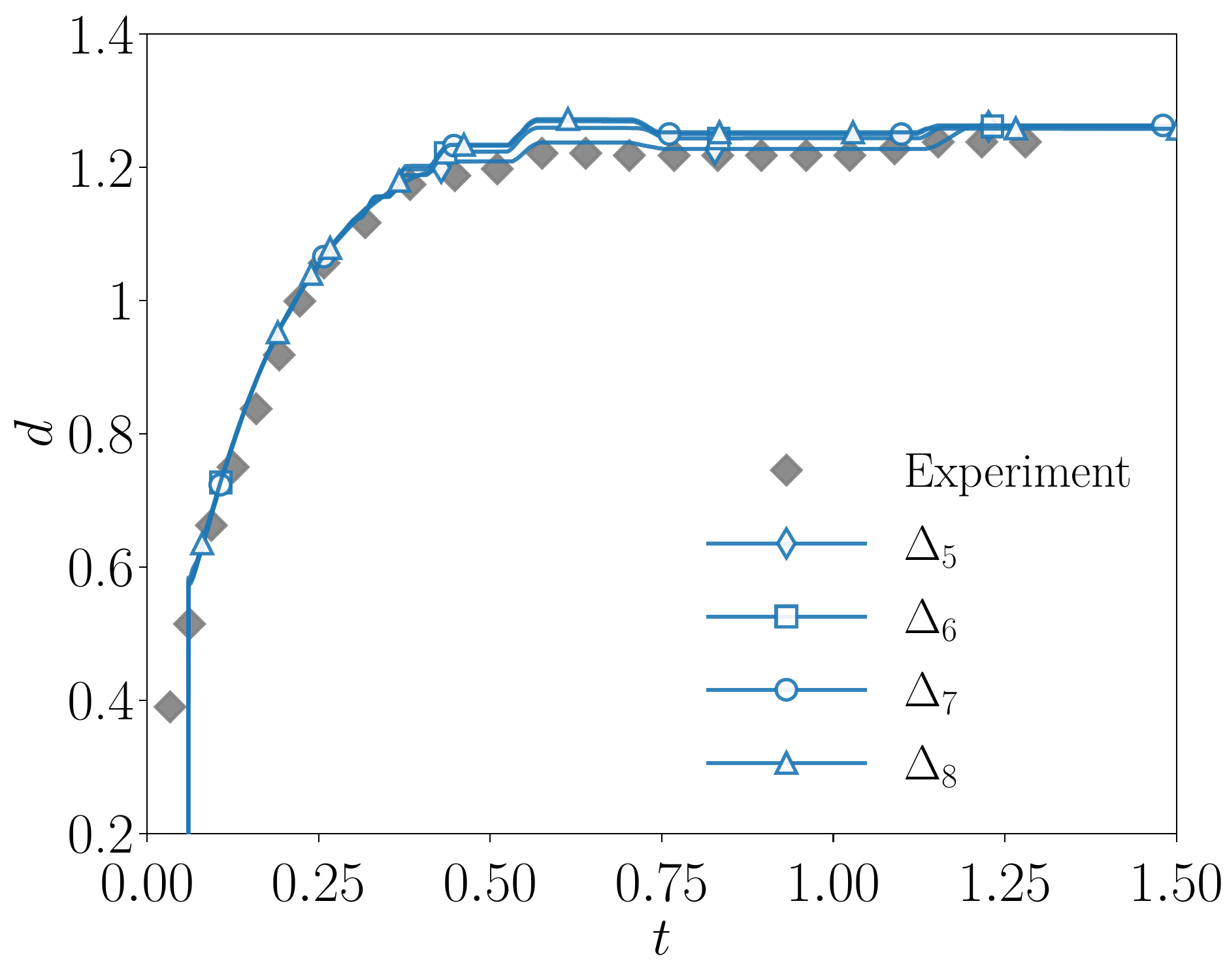}
\caption{Comparison of the temporal evolution of the experimentally obtained \citep{roisman2008drop} contact-line position with the present simulations at different grid resolutions without the proposed subgrid model. This liquid-gas system shows grid-independence even without a grid-dependent contact-angle model.}
\label{fig:roisman_spreading}
\end{figure}

\begin{figure}[htpb]
\centering
\includegraphics[width=0.9\columnwidth]{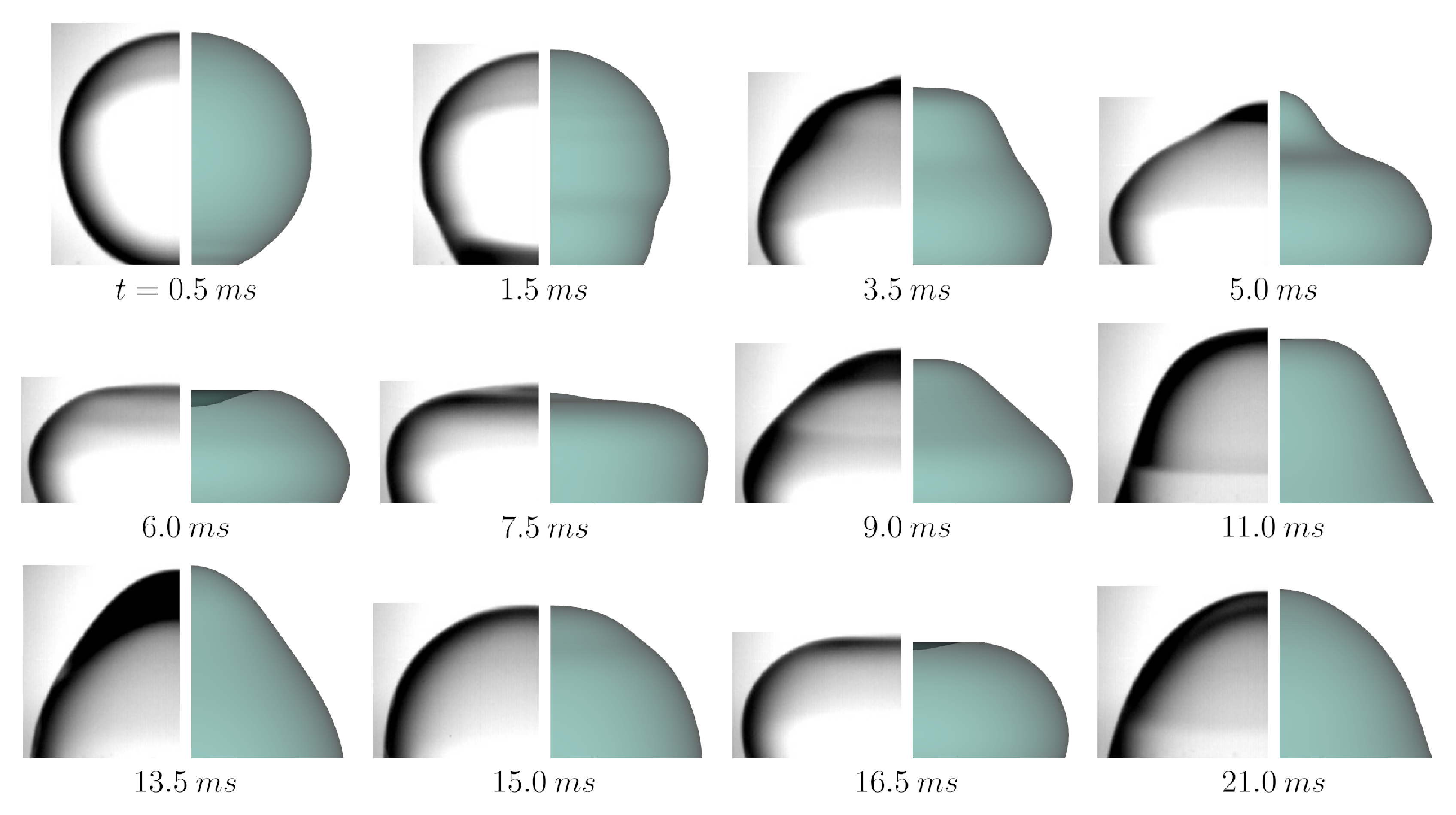}
\caption{Comparison of the droplet shapes at different time instants obtained experimentally \citep{roisman2008drop} against the present numerical simulations without the proposed subgrid model.}
\label{fig:roisman_shapes}
\end{figure}

\begin{figure}[htp]
    \centering
    \includegraphics[width=13cm]{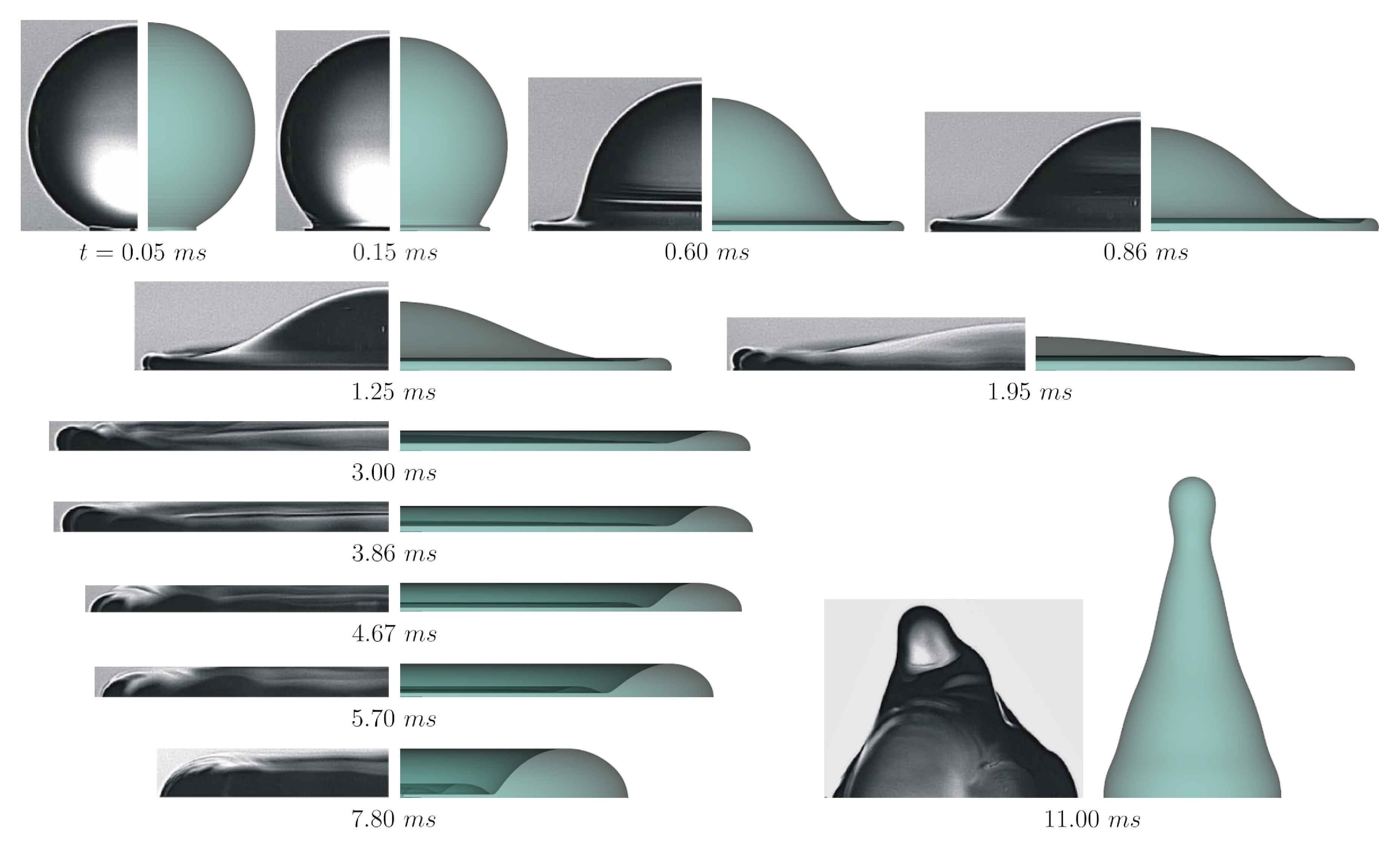}
    \caption{Comparison of droplet shapes obtained from the present numerical simulations with the experimentally observed droplet shapes of \citet{vsikalo2005dynamic} at different time instances.}
    \label{fig:sikaloshapes}
\end{figure}

The contact-line position obtained from our simulations match well with that given by Roisman et al.~\citet{roisman2008drop}, as shown in Fig.~\ref{fig:roisman_spreading}. Additionally, Fig.~\ref{fig:roisman_shapes} shows a good agreement between the droplet shapes of our simulation and those from the experiments at various time instants. Moreover, the contact line position appears to be grid-independent with the maximum error between the results of grids $\Delta_6$ and $\Delta_{8}$ less than $0.1\%$. A formal grid-convergence test is often not supplied in the literature for such simulations, possibly for this reason. Similarly, another droplet impact case from the literature~\citep{vsikalo2005dynamic} shows a good match with the experimental results, as shown in Fig.~\ref{fig:sikaloshapes}.

Now, we move our attention to another droplet spreading case with a different set of parameters where the grid-dependence is reported to be significant \citep{afkhami2009mesh}.

\subsubsection{Case B: Liquid-liquid system}
\label{sec:caseB}
\begin{figure}
  \centerline{%
    \begin{minipage}[t]{.46875\columnwidth}
      \centering
      {\small (a)}\\[1mm]
      \includegraphics[width=\linewidth]{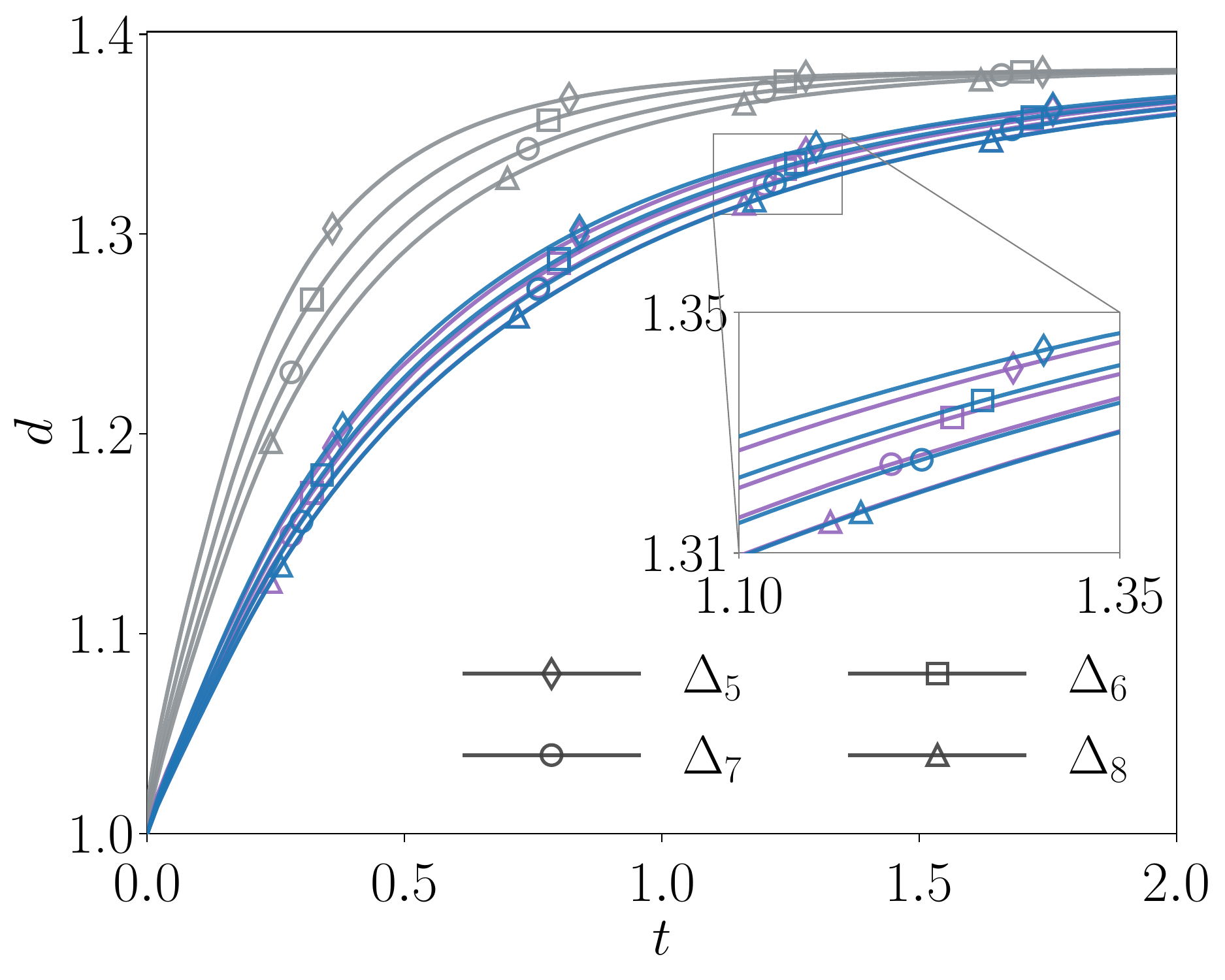}
    \end{minipage}
    \hspace{8mm}
    \begin{minipage}[t]{.46875\columnwidth}
      \centering
      {\small (b)}\\[1mm]
      \includegraphics[width=\linewidth]{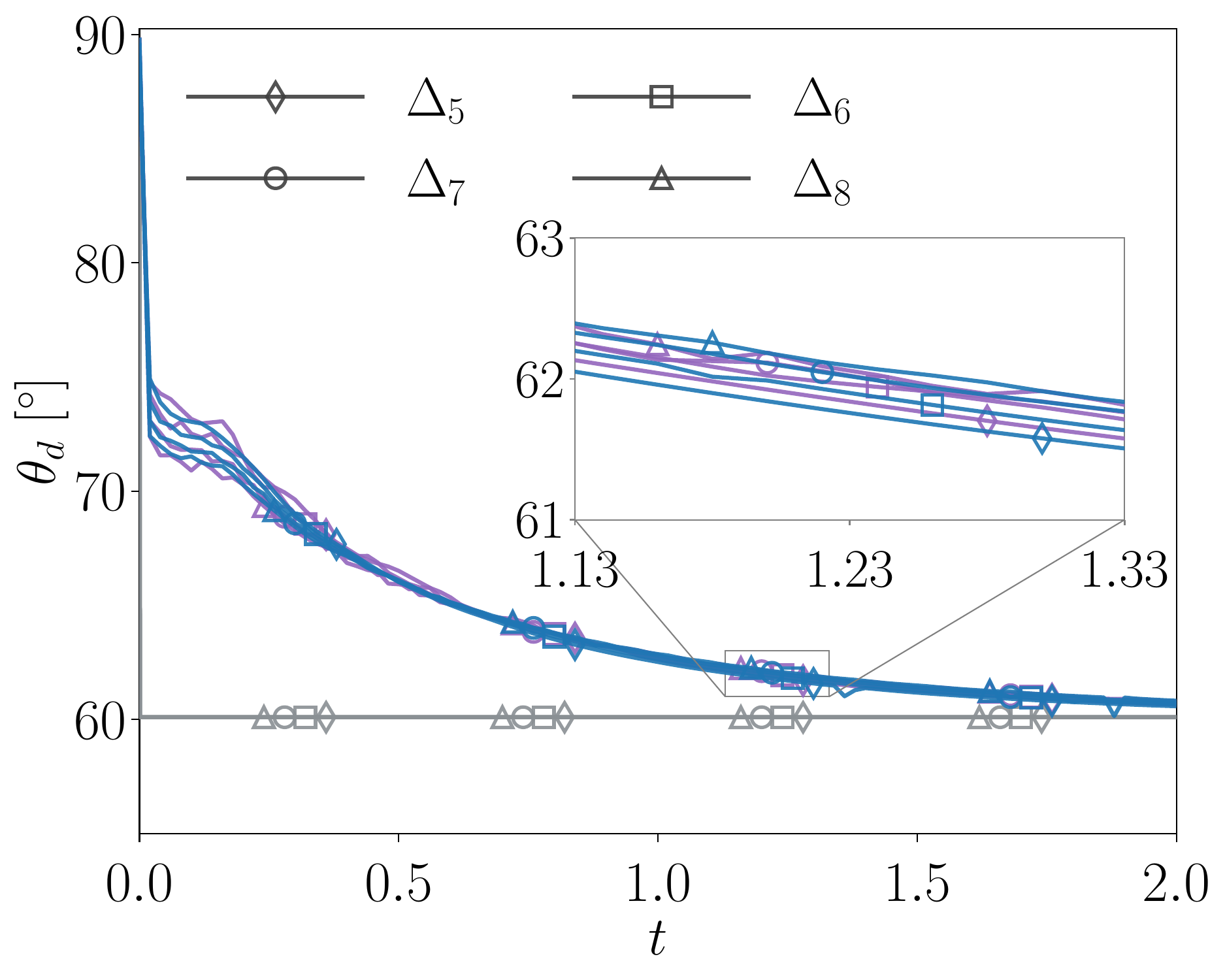}
    \end{minipage}
  }
  \vspace{2mm}
  \centerline{%
    \begin{minipage}[t]{.46875\columnwidth}
      \centering
      {\small (c)}\\[1mm]
      \includegraphics[width=\linewidth]{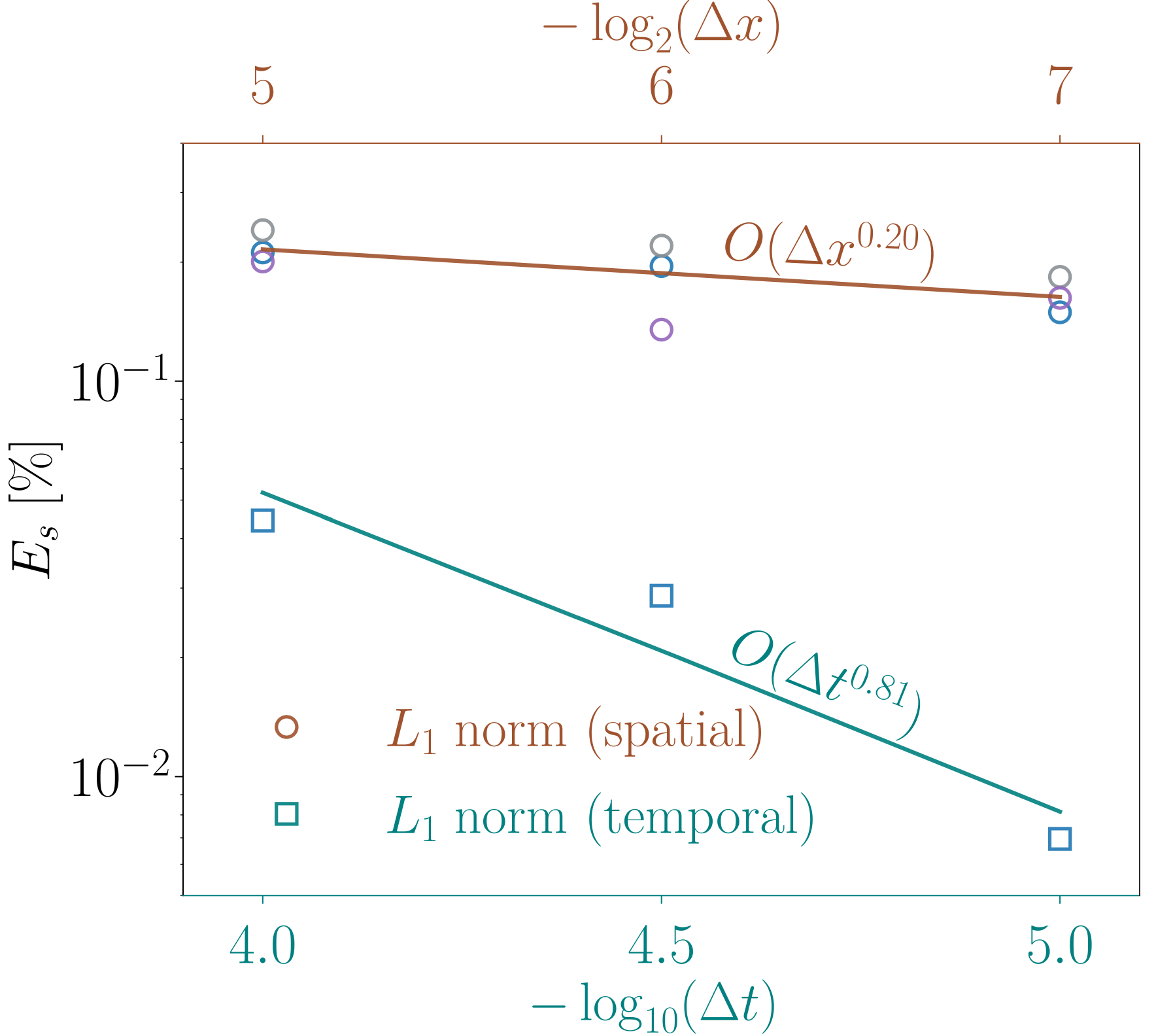}
    \end{minipage}
    \hspace{8mm}
    \begin{minipage}[t]{.46875\columnwidth}
      \centering
      {\small (d)}\\[3.5mm]
      \includegraphics[width=\linewidth]{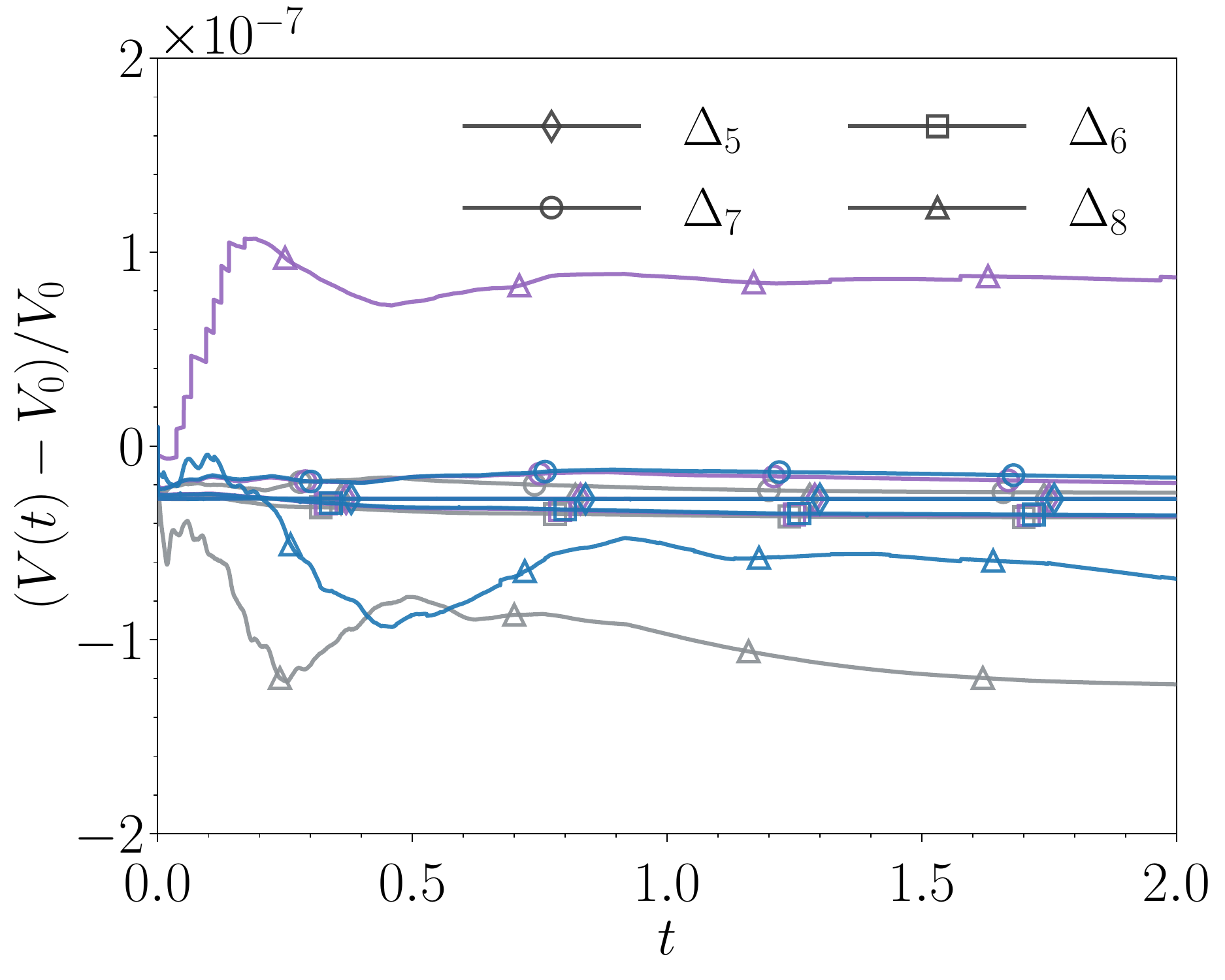}
    \end{minipage}
  }

  \caption{Plots obtained from the present simulations of the liquid-liquid system, case `B' (Sec. \ref{sec:caseB}) without the proposed model, showing (a) the temporal evolution of spreading diameter ($d$) for different grid-sizes, (b) the temporal evolution of dynamic contact angle ($\theta_d$) for different grid-sizes, (c) the variation of droplet-shape error $E_s^{L_1}$ with grid-size and time-step, and (d) the variation of the relative droplet volume with time for different grid-sizes. In panels (a), (b), and (d), the Kistler model with the contact-line speed model proposed in the present work, the Kistler model with the contact-line speed model proposed by \cite{roisman2008drop}, and the quasi-static model are represented by \textcolor[rgb]{0.1216,0.4667,0.7059}{\rule[0.5ex]{2.4em}{1pt}}, \textcolor[rgb]{0.502,0,0.502}{\rule[0.5ex]{2.4em}{1pt}}, and \textcolor{gray}{\rule[0.5ex]{2.4em}{1pt}}, respectively. In panel (c), symbols denote the droplet-shape error for different $\Delta x$ (top axis) and $\Delta t$ (bottom axis), evaluated between consecutive resolutions using the finer resolution as the reference.
}
    \label{fig:afkh_compare}
\end{figure}

We perform the simulations in an axisymmetric domain using the same initial conditions and boundary conditions as those employed by \citet{afkhami2009mesh}. The corresponding non-dimensional parameters, with the initial downward velocity as the velocity scale, are $\rho_r=1$, $\mu_r=1$, $Re=2$ and $We=0.067$. The equilibrium contact angle is set to $60^\circ$ to ensure that the contact line remains advancing throughout the simulation, thereby eliminating the effects of contact-angle hysteresis and requiring only an advancing contact angle model. The simulations are performed on four adaptive grids with the finest grid sizes $\Delta_i=D_0/2^{i}$, where $i=5,6,7,$ and $8$. The results of two contact angle models, namely Kistler's \citep{kistler1993hydrodynamics} contact angle model and fixed contact angle model, and two contact-line velocity models, namely the extrapolated interface velocity method proposed by \citet{roisman2008drop} and the present method as discussed in \ref{subsubsec:theta_d}, are compared by simulating three different cases (see Fig.~\ref{fig:afkh_compare}) which are as follows: a) Fixed contact angle, b) Kistler's contact angle model with extrapolated contact line velocity as given in \citet{roisman2008drop}, and c) Kistler's contact angle model with the present method of contact line speed calculation. 

Contact-line position exhibits noticeable dependence on the grid resolution for all three cases considered, as shown in Fig.~\ref{fig:afkh_compare}(a). This behavior is a typical manifestation of grid dependence and is consistent with the observations of \citet{afkhami2009mesh}. However, the results obtained with speed-dependent contact-angle models exhibit a closer clustering of solutions across grid sizes as compared to that for the fixed-contact-angle model, indicating a noticeable reduction in grid-sensitivity-induced errors. Nevertheless, the errors are considerably larger as compared to those for the gas-liquid cases considered in the previous section. The results also indicate that both approaches of contact-line speed calculation yield nearly identical predictions. Furthermore, Fig.~\ref{fig:afkh_compare}(b) demonstrates that the temporal order of convergence for the present method is approximately $1$, while the spatial order of convergence for all three cases is observed to be nearly zero, which shows that the method is non-convergent. Fig.~\ref{fig:afkh_compare}(c) shows that the dynamic contact angle varies negligibly with grid size for all three cases, except for the early-time deviations in the case of speed-dependent contact angle models. To rule out the errors stemming from inaccurate imposition of mass conservation, Fig.~\ref{fig:afkh_compare}(d) shows the maximum relative error in droplet volume to be less than $\sim 10^{-7}$ for all grid sizes. Altogether, these results show that the grid-dependence is due to an improper incorporation of the empirical contact angle models in the numerical method, which we investigate next.

Additionally, it should be noted that the error between the results obtained from different grids is considerably smaller for gas-liquid cases than that for the liquid-liquid cases. Therefore, the present numerical method, and the ones already existing \citep{vsikalo2005dynamic,roisman2008drop} may be sufficient for studying moving gas-liquid contact-line flows. A similar conclusion was reached for the gas-liquid systems by \cite{legendre2015comparison} while applying various dynamic contact angle models.

\section{Grid-dependence in contact-line flow problems} 

\label{sec:thetad}
\begin{figure}
    \centering
      \includegraphics[width=8.5cm]{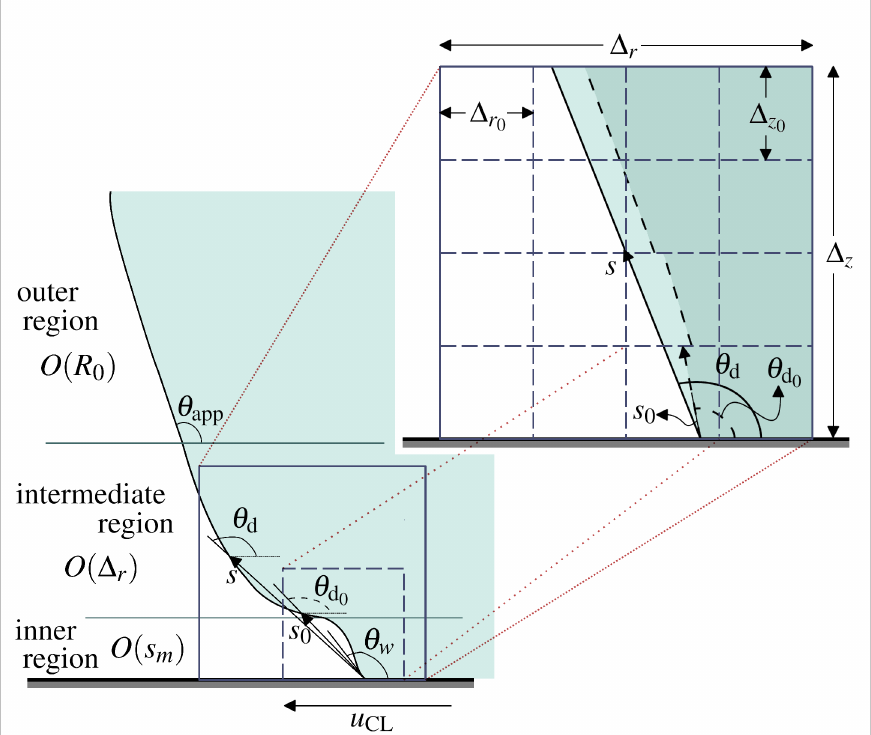}
    \caption{Multiscale nature of the interface in the region encompassing the moving contact-line according to \cite{cox1986dynamics}, showing the inner, intermediate, and outer regions, with the inset showing the contact-line cells in the intermediate region at two different, coarse ($\Delta_r,\Delta_z$) and fine ($\Delta_{r_0},\Delta_{z_0}$), grid resolutions, illustrating the imposed angles $\theta_d$ and $\theta_{d0}$ at their respective cell-centres.
}
\label{fig:coxcell}
\end{figure}

It has been observed experimentally \citep{mues1989observation,rame1991identifying} and established theoretically by the hydrodynamic analysis of \cite{cox1986dynamics} that, in regions beyond the microscopic region—namely the intermediate and the outer (macroscopic) regions (see Fig. \ref{fig:coxcell}) — the observed contact angle varies with the distance from the contact line ($s$) because of a competition between the surface tension and viscous forces. Mathematically, in the intermediate region, the matched-asymptotic relation as given in \citet{cox1986dynamics} is written as
\begin{equation}
G(\theta(s))
=
G(\theta_w)
+
Ca_{\mathrm{CL}}\ln\!\left(\frac{s}{\lambda}\right)
+
Ca_{\mathrm{CL}}
\left[
\frac{Q_i}{f(\theta_w,\mu_r)}
\right] + \mathcal{{O}}(Ca_{\mathrm{CL}}^2),
\label{eq:cox_full}
\end{equation}
where the Cox's hydrodynamic function is defined as
\begin{equation}
G(\theta)
=
\int_{0}^{\theta}
\frac{d\phi}{f(\phi,\mu_r)},
\label{eq:G_def}
\end{equation}
with
\begin{equation}
f(\phi, \mu_r) =
\frac{
2 \sin\phi \left\{ \mu_r^2 (\phi^2 - \sin^2\phi) + 2\mu_r [\phi(\pi - \phi) + \sin^2\phi] + [(\pi - \phi)^2 - \sin^2\phi] \right\}
}{
\mu_r (\phi^2 - \sin^2\phi) [(\pi - \phi) + \cos\phi \sin\phi] + [(\pi - \phi)^2 - \sin^2\phi] (\phi - \cos\phi \sin\phi)
}.
\end{equation}    

wherein $\theta(s)$ is the angle that the position vector to the interface, $\vec{s}$, makes with the solid surface, and $\theta_w$ is the microscopic contact angle at the wall, i.e., $s\to 0$, as shown in Fig. \ref{fig:coxcell}. Contact-line capillary number, $Ca_{\mathrm{CL}}$, is also defined at the wall, and $\lambda$ is a characteristic length scale in the inner region. 
$Q_i$ is a constant arising from the matching between the inner and intermediate asymptotic regions. Eq. \eqref{eq:cox_full} can be conveniently rewritten by introducing a `microscopic length scale' $s_m$ \citep{afkhami2018} as follows: 
\begin{equation}
    G(\theta(s)) = G(\theta_w)+Ca_{\mathrm{CL}} \ln\!\left(\frac{s}{s_m}\right)
    \label{eq:COX_final}
\end{equation}

\begin{figure}
  \centerline{%
    \begin{minipage}[t]{.46875\columnwidth}
      \centering
      {\small (a)}\\[1mm]
      \includegraphics[width=\linewidth]{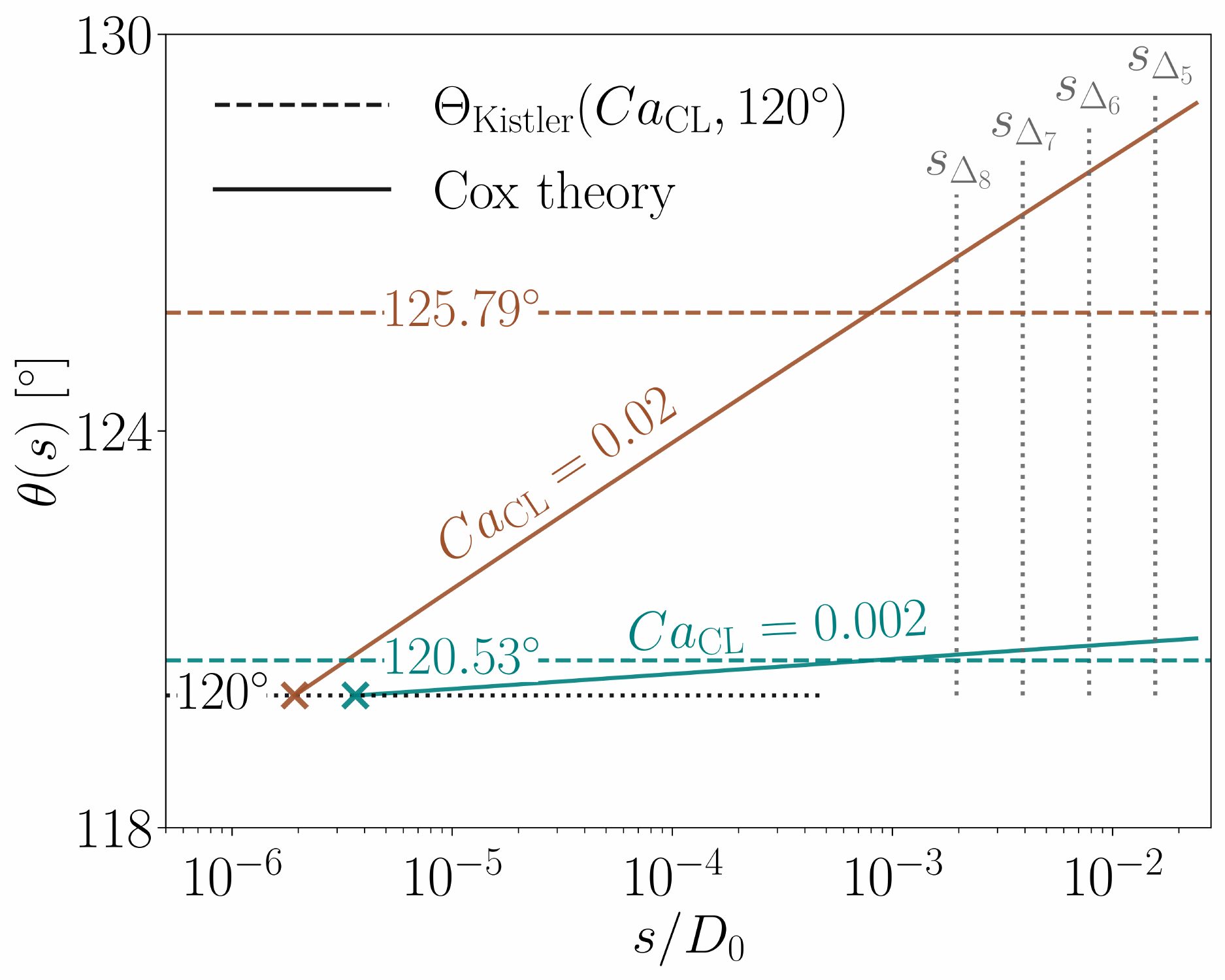}
    \end{minipage}
    \hspace{8mm}
    \begin{minipage}[t]{.46875\columnwidth}
      \centering
      {\small (b)}\\[1mm]
      \includegraphics[width=\linewidth]{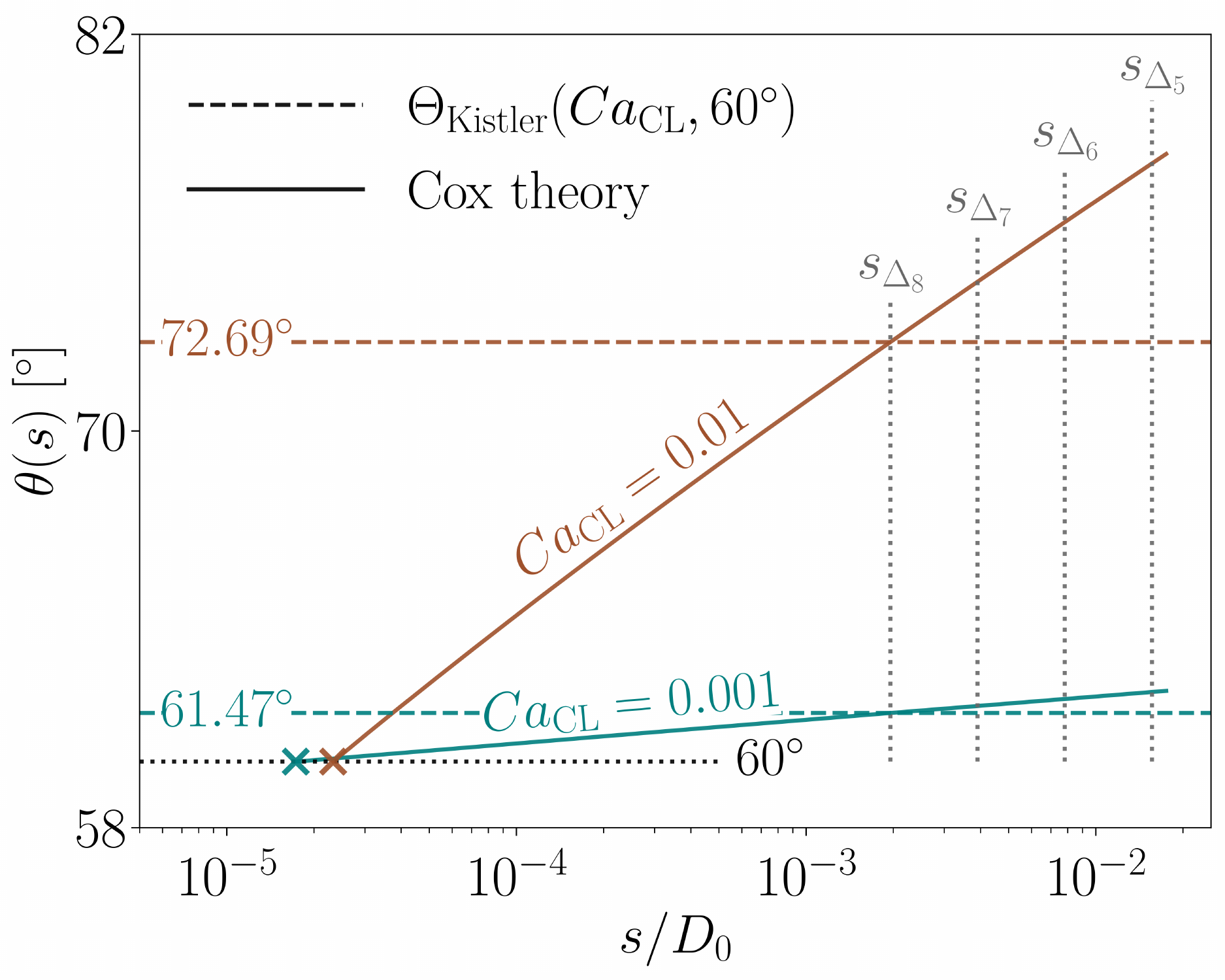}
    \end{minipage}
  }

  \caption{Comparison between the contact angles obtained from the Cox thoery
  (Eq.~\eqref{eq:COX_final}), $\theta(s)$, and from the Kistler dynamic contact angle
  (Table~\ref{tab:models}) for (a) case `A' where $Ca_{\mathrm{CL}}$ remains between $=0.002$ and $0.02$, and (b) case `B' where $Ca_{\mathrm{CL}}$ remains between $=0.001$ and $0.01$, for most of their dynamics. The plots for the higher and the lower $Ca_{\mathrm{CL}}$ are represented by \textcolor[rgb]{0.6275,0.3216,0.1765}{\rule[0.5ex]{2.4em}{1pt}} and \textcolor[rgb]{0,0.5,0.5}{\rule[0.5ex]{2.4em}{1pt}}, respectively, in each of the panels.}
\label{fig:cox_angle}
\end{figure}

Fig. \ref{fig:cox_angle} shows the variation of $\theta$ with the distance $s$ from the contact line as given by Eq.~\eqref{eq:COX_final} for both cases considered in the previous section. The contact angles measured at a distance $s$ for two values of $Ca_{\mathrm{CL}}$ encompassing most of the dynamics, i.e., $0.002$ and $0.02$ for case `A', and $0.001$ and $0.01$ for case `B', are shown in this figure. For case `A', the deviation in contact angle at different distances from the contact line is smaller than that for case `B'. In addition, the empirical contact angle used in the simulation (from the Kistler model) matches the angle predicted by Cox, which leads to a good match between the simulation and the experiment. For Case `B', the empirical contact angle approximately matches the angle predicted by Cox for the grid size $\Delta_8$, whereas the angle predicted by Cox at other grids is significantly different from the numerically imposed empirical contact angle (the difference being about $5.5^\circ$ for $\Delta_7$, $8.0^\circ$ for $\Delta_6$, and  $10.5^\circ$ for $\Delta_5$ for $Ca_{\mathrm{CL}} = 0.01$). Therefore, the numerical contact angle $\theta_d$ should vary with the grid size, as also concluded by \citet{afkhami2009mesh,afkhami2018}.  

The empirical contact angle models capture the interface slope and position at a distance, generally a few micrometers, away from the contact line, which depends on the resolution of the measuring instrument \citep{kistler1993hydrodynamics,spelt2005level}, henceforth called the observation scale, $s_0$, of the model. Simulations can be made to match the experiments by selecting a grid size at the observation scale so that the imposed empirical contact angle, $\theta_d$ satisfies the Cox formula (\eqref{eq:COX_final}). The fortuitous match with the experimental results in the simulation of case `A' may be attributed to the fact that the grid size is approximately the same as $s_0$, and the fact that the variation of contact angle with $s$ is less, so that a mesh refinement does not cause the angle to deviate noticeably.

It may not always be possible to choose a grid-size that matches $s_0$. Moreover, to achieve grid-independence, in general, we need to know the value of $\theta_w$ and $s_m$ a priori, which is difficult to ascertain from experiments and molecular simulations \citep{afkhami2018}. \citet{10.1021/la00043a013} coupled the molecular-kinetic equation of \citet{BLAKE1969421}, which relates $\theta_w$ and $u_{CL}$, with Cox's model as given in Eq.~\eqref{eq:COX_final}, for calculating $\theta(s)$; but the value of $s/s_m$ was taken as a fitting parameter \citep{ren2010continuum}, rendering predictive simulations impossible. Ideally, the numerical method should yield grid-independent results without the need of any problem-specific fine-tuning of the parameters.      

The difference in the observed displacement of the contact line at the observation scale, with respect to the actual, microscopic, contact line displacement, is limited to the intermediate scale. This difference in the two displacements occurs only for accelerating contact lines due to a temporal variation of the interface slope. The effect of this displacement difference on the overall dynamics at the macroscopic length scale is insignificant in contrast to the effect on the interface slope due to viscous bending. Therefore, for macroscopic flows, the contact line position may be assumed to be independent of the observation scale as long as the latter is in the intermediate region. This implies that the observed capillary number may be assumed to be equal to $Ca_{\mathrm{CL}}$. Thus, the indicator of grid-independence of the numerical method is the grid-independence of the contact line position instead of that of the contact angle, in contrast to the observations in Figs.~\ref{fig:afkh_compare}(a) and \ref{fig:afkh_compare}(b).

\section{Coupling of empirical contact-angle models with Cox’s hydrodynamic theory}
\label{sec:coupling_cox}
As discussed in the previous section, empirical dynamic contact-angle models, $\Theta(Ca_{\mathrm{CL}},\theta_e)$ $(\equiv\theta_0)$, prescribe the dynamic contact angle at a finite observation scale $s_0$ associated with the experimental resolution and therefore cannot directly predict the contact angle at an arbitrary distance $s$, corresponding to the grid-size. To extend these models to arbitrary distances within the intermediate region, we couple them with the Cox-Voinov theory. In the limiting case, $Ca_{\mathrm{CL}}=0$, empirical dynamic contact-angle models recover the equilibrium contact angle, $\Theta(0,\theta_e)=\theta_e$, whereas Cox's theory predicts the contact angle at any distance $s$ to be the microscopic wall angle, $\theta_w$, which cannot be assumed to be equal to $\theta_e$. Past attempts to make the wetting simulations grid-independent have assumed $\theta_w$ to be constant equal to $\theta_e$, and $s_m$ to be a constant \citep{afkhami2009mesh,legendre2015comparison}, which may not be the case. In contrast, we do not place any such restrictions on $\theta_w$ and $s_m$ in favour of a more general model. 

As the empirically observed contact angle must also satisfy Cox’s hydrodynamic relation, Eq.~\eqref{eq:COX_final} at $s_0$, therefore, 
\begin{equation}
G\!\left(\theta(s_0)\right)
=
G\!\left(\theta_0\right)
=
G(\theta_e)
+
Ca_{\mathrm{CL}}
\ln\!\left(\frac{s_0}{s_m}\right).
\label{eq:taylor_right1}
\end{equation}
By combining this with Eq.~\eqref{eq:COX_final}, which is given for some arbitrary $s$, we get 
\begin{equation}
G\!\left(\theta(s)\right)=
G(\theta_0)
+
Ca_{\mathrm{CL}}
\ln\!\left(\frac{s}{s_0}\right),
\label{eq:implicit_final}
\end{equation}
which is entirely expressed in terms of the empirical model's observation scale, $s_0$, and observed contact angle $\theta_0$, thereby eliminating the microscopic length scale $s_\mathrm{m}$ and wall angle, $\theta_w$. Therefore, once $s_0$ is known along with the empirical model, the effect of viscous bending can be integrated in numerical simulations according to the following steps. First the angle $\theta_0$ is obtained from the empirical relation $\Theta(Ca_{\mathrm{CL}},\theta_e)$, and then applying Eq.~\eqref{eq:implicit_final} gives the scale-dependent correction for the angle to be imposed at the grid level; that is, for a grid size $\Delta$, the contact angle associated with the locally reconstructed PLIC interface is imposed at $\Delta/2$, which may be approximated as $s$.

To incorporate the above two-step procedure directly into the dynamic empirical model $\Theta$, we assume a scale-dependent dimensionless number $\chi(Ca_{\mathrm{CL}}, s)$ in place of $Ca_{\mathrm{CL}}$ in the `original' empirical model $\Theta(Ca_{\mathrm{CL}}, \theta_e)$,  such that, the `modified' model is given by
\begin{equation}
    \Theta\!\left(\chi(Ca_{\mathrm{CL}},s),\theta_e\right)=\theta(s),
    \label{eq:effective_ca}
\end{equation}
with the limiting condition
\begin{equation}
    \chi(0,s)=0,
\end{equation}
which ensures
\begin{equation}
    \Theta\!\left(\chi(0,s),\theta_e\right)=\Theta(0,\theta_e)=\theta_e,
\end{equation}
thereby recovering the static limit of Cox's theory. Also, at the observation scale of dynamic empirical models, $s_0$, it should satisfy,
\begin{equation}            
\chi(Ca_{\mathrm{CL}},s_0) =Ca_{\mathrm{CL}},
    \label{eq:chi_s0}
\end{equation}
so that the experimentally observed contact angle, $\theta_0  = \Theta\!\left(\chi(Ca_{\mathrm{CL}},s_0),\theta_e\right)=\Theta\!\left(Ca_{\mathrm{CL}},\theta_e\right).$
For a given $\theta_e$ and $Ca_{\mathrm{CL}}$, $\chi(s)$ at an arbitrary distance $s$, in the intermediate region, can be given by, 
\begin{equation}
\chi(s) = \chi(s_0) + \delta \chi= Ca_{\mathrm{CL}} + \delta \chi,
\label{eq:chi(s)}
\end{equation}
and the Taylor expansion of the composite function $G(\Theta(\chi(s)))$ about $\chi(s_0)$ is then given by,
\begin{align}
G(\Theta(\chi(s)))
&\approx
G(\theta_0)
+
\left.
\frac{dG}{d\Theta}
\frac{d\Theta}{d\chi}
\right|_{\chi=Ca_{\mathrm{CL}}}
\delta\chi
+
\frac{1}{2}
\left.
\frac{d}{d\chi}
\left(
\frac{dG}{d\Theta}
\frac{d\Theta}{d\chi}
\right)
\right|_{\chi=Ca_{\mathrm{CL}}}
(\delta\chi)^2
+
O((\delta\chi)^3).
\label{eq:taylor_second_order}
\end{align}
For $|\delta \chi|\ll 1$, higher-order terms in the Eq.~\eqref{eq:taylor_second_order} can be neglected, thereby yielding
\begin{equation}
G(\Theta(\chi(s)))
\approx
G(\theta_0)
+
\left.
\frac{dG}{d\Theta}
\right|_{\Theta(\chi(s_0))}
\left.
\frac{d\Theta}{d\chi}
\right|_{\chi=Ca_{\mathrm{CL}}}
\delta \chi.
\label{eq:taylor_left}
\end{equation}
From Eqs.~\eqref{eq:implicit_final}, \eqref{eq:effective_ca} and \eqref{eq:taylor_left}, we get
\begin{equation}
\left.
\frac{dG}{d\Theta}
\right|_{\Theta=\theta_0}
\left.
\frac{d\Theta}{d\chi}
\right|_{\chi=Ca_{\mathrm{CL}}}
\delta \chi
=
Ca_{\mathrm{CL}}
\ln\!\left(\frac{s}{s_0}\right).
\label{eq:linear_balance}
\end{equation}
This leads to
\begin{equation}
\delta \chi
=
\frac{Ca_{\mathrm{CL}}}
{
\left.
\dfrac{dG}{d\Theta}
\right|_{\Theta=\theta_0}
\left.
\dfrac{d\Theta}{d\chi}
\right|_{\chi=Ca_{\mathrm{CL}}}
}
\ln\!\left(\frac{s}{s_0}\right).
\label{eq:deltaCa_general}
\end{equation}
Since $\chi(Ca_{\mathrm{CL}},s)$ is a modification to $Ca_{\mathrm{CL}}$ in the empirical relation $\Theta$, without altering the functional form of empirical relation itself,
$\left.\dfrac{d\Theta}{d\chi}\right|_{\chi=Ca_{\mathrm{CL}}}
=
\left.\dfrac{d\Theta}{dCa_{\mathrm{CL}}}\right.$
and from Eq.~\eqref{eq:effective_ca}, we can write,
$\left.\dfrac{dG}{d\Theta}\right|_{\Theta=\theta_0}
=
\left.\dfrac{dG}{d\theta}\right|_{\theta=\theta_0}$,
and from Eq.~\eqref{eq:G_def}, we obtain,
\begin{equation}
\delta \chi
=
Ca_{\mathrm{CL}}
\frac{f(\theta_0,\mu_r)}
{\left.\dfrac{d\Theta}{dCa_{\mathrm{CL}}}\right.}
\ln\!\left(\frac{s}{s_0}\right).
\label{eq:deltaCa}
\end{equation}
Since $\theta_0  =\Theta\!\left(Ca_{\mathrm{CL}},\theta_e\right)$, Eqs.~\eqref{eq:chi(s)} and \eqref{eq:deltaCa} lead to
\begin{equation}
\chi(s)
=
Ca_{\mathrm{CL}}
\left[
1
+
\frac{f(\Theta(Ca_{\mathrm{CL}},\theta_e),\mu_r)}
{\left.\dfrac{d\Theta}{dCa_{\mathrm{CL}}}\right.}
\ln\!\left(\frac{s}{s_0}\right)
\right].
\label{eq:Ca_linear}
\end{equation}
For convenience, we define $\zeta(Ca_{\mathrm{CL}})$ as
\begin{equation}
\zeta
=
\frac{f(\Theta(Ca_{\mathrm{CL}},\theta_e),\mu_r)}
{\left.\dfrac{d\Theta}{dCa_{\mathrm{CL}}}\right.},
\label{eq:zeta_def}
\end{equation}
which can be directly calculated from a given empirical model. Therefore, for a known observation scale of the empirical model,
\begin{equation}
\chi(Ca_{\mathrm{CL}},s)
=
Ca_{\mathrm{CL}}
\left[
1
+
\zeta\,
\ln\!\left(\frac{s}{s_0}\right)
\right].
\label{eq:zeta_final}  
\end{equation}

\begin{figure}
  \centerline{%
    \begin{minipage}[t]{.46875\columnwidth}
      \centering
      \small{(a)}\\[1mm]
      \includegraphics[width=\linewidth]{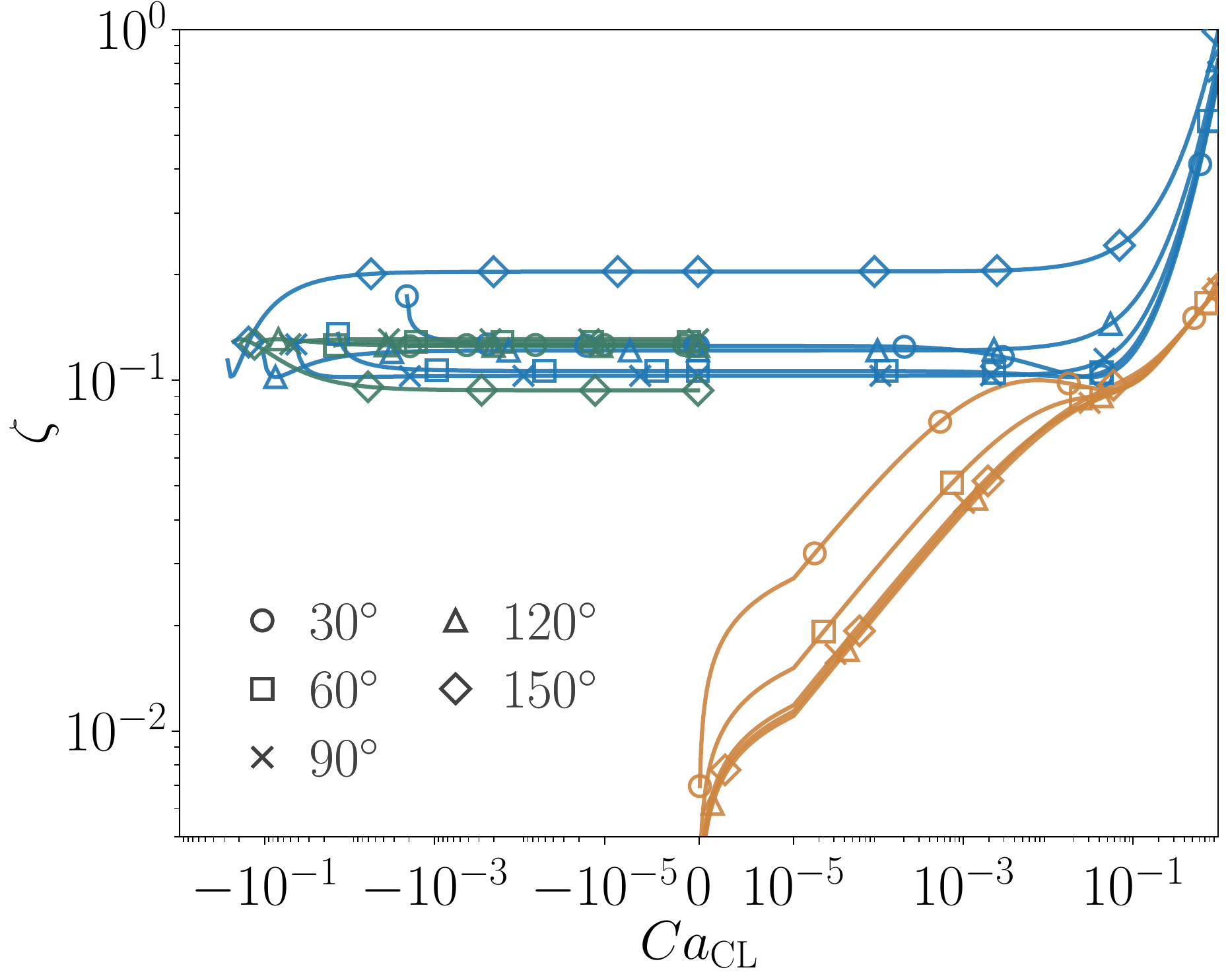}
    \end{minipage}
    \hspace{8mm}
    \begin{minipage}[t]{.46875\columnwidth}
      \centering
      \small{(b)}\\[1mm]
      \includegraphics[width=\linewidth]{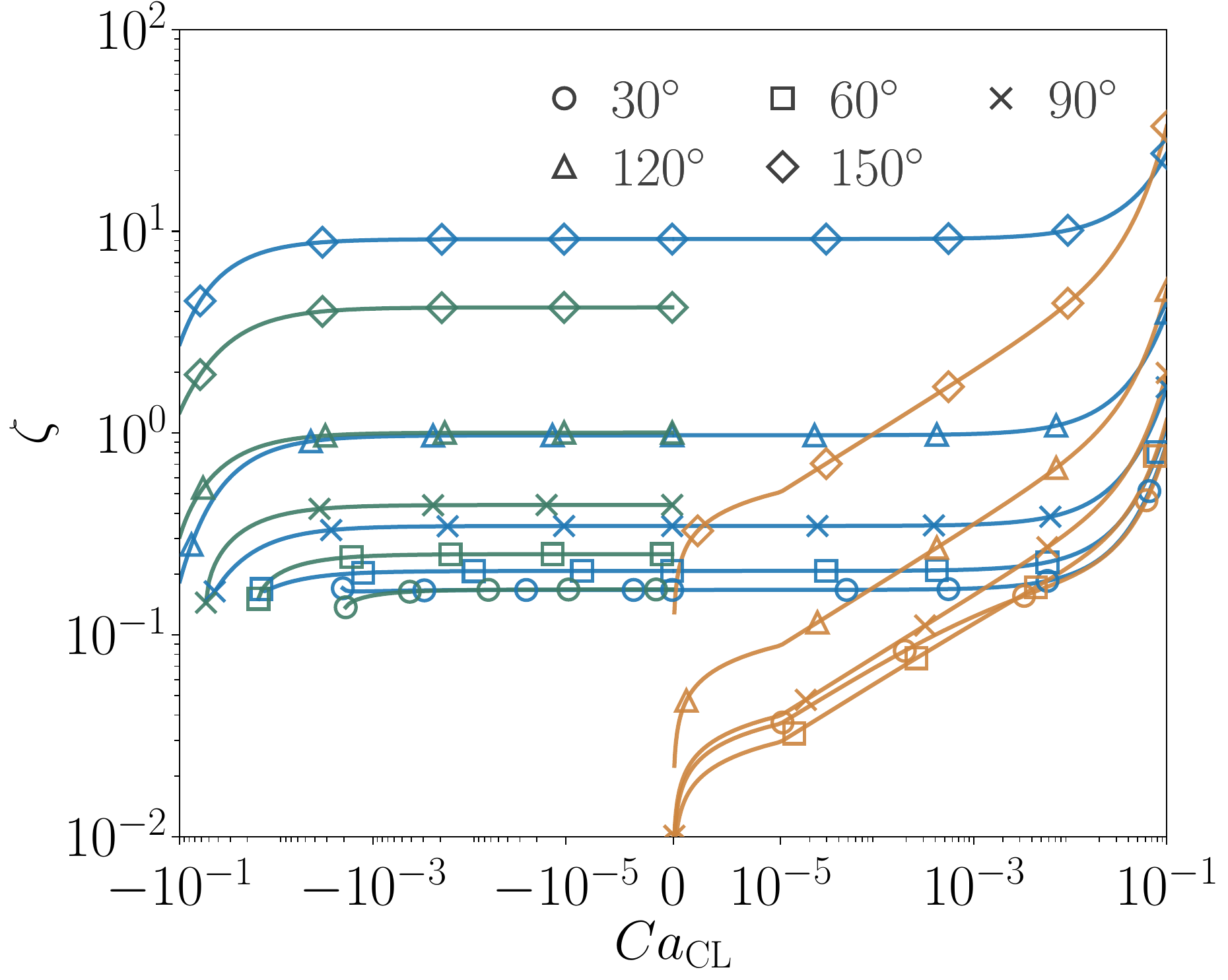}
    \end{minipage}
  }

  \caption{Variation of $\zeta$ with the contact-line capillary number, $Ca_{\mathrm{CL}}$, for different equilibrium contact angles with (a) $\mu_r=0$ and (b) $\mu_r=1$. The Kistler, Jiang, and Tanner dynamic contact-angle models are represented by \textcolor[rgb]{0.1216,0.4667,0.7059}{\rule[0.5ex]{2.4em}{1pt}}, \textcolor[rgb]{0.8039,0.5216,0.2471}{\rule[0.5ex]{2.4em}{1pt}} and \textcolor[rgb]{0.2471,0.4902,0.4078}{\rule[0.5ex]{2.4em}{1pt}}, respectively.
}
  \label{fig:zeta_angle}
\end{figure}

We first examine the behavior of the parameter $\zeta$ over a range of capillary numbers and contact angles. Fig.~\ref{fig:zeta_angle} shows that $\zeta$, as given by Eq.~\eqref{eq:zeta_def}, remains approximately constant for $Ca_{\mathrm{CL}}<0.1$ over the entire range of contact angles considered for the `Kistler' and `Tanner' models. In contrast, for the `Jiang' model, $\zeta$ increases with $Ca_{\mathrm{CL}}$ at all contact angles considered.

Therefore, before testing the numerical simulations with the value of $\zeta$ obtained from Eq.~\ref{eq:zeta_def} for grid-independence, we run simulations with different, constant, values of $\zeta$ for case `B' with the modified Kistler's contact angle model for $\Delta_5$, $\Delta_6$, and $\Delta_7$ grid sizes. For these simulations, we choose $\Delta_8$ as $s_0$, which is evident from Fig. \ref{fig:cox_angle}(b), therefore making it the reference grid-size.

\begin{figure}
  \centerline{%
    \begin{minipage}[t]{.46875\columnwidth}
      \centering
      {\small (a)}\\[1mm]
      \includegraphics[width=\linewidth]{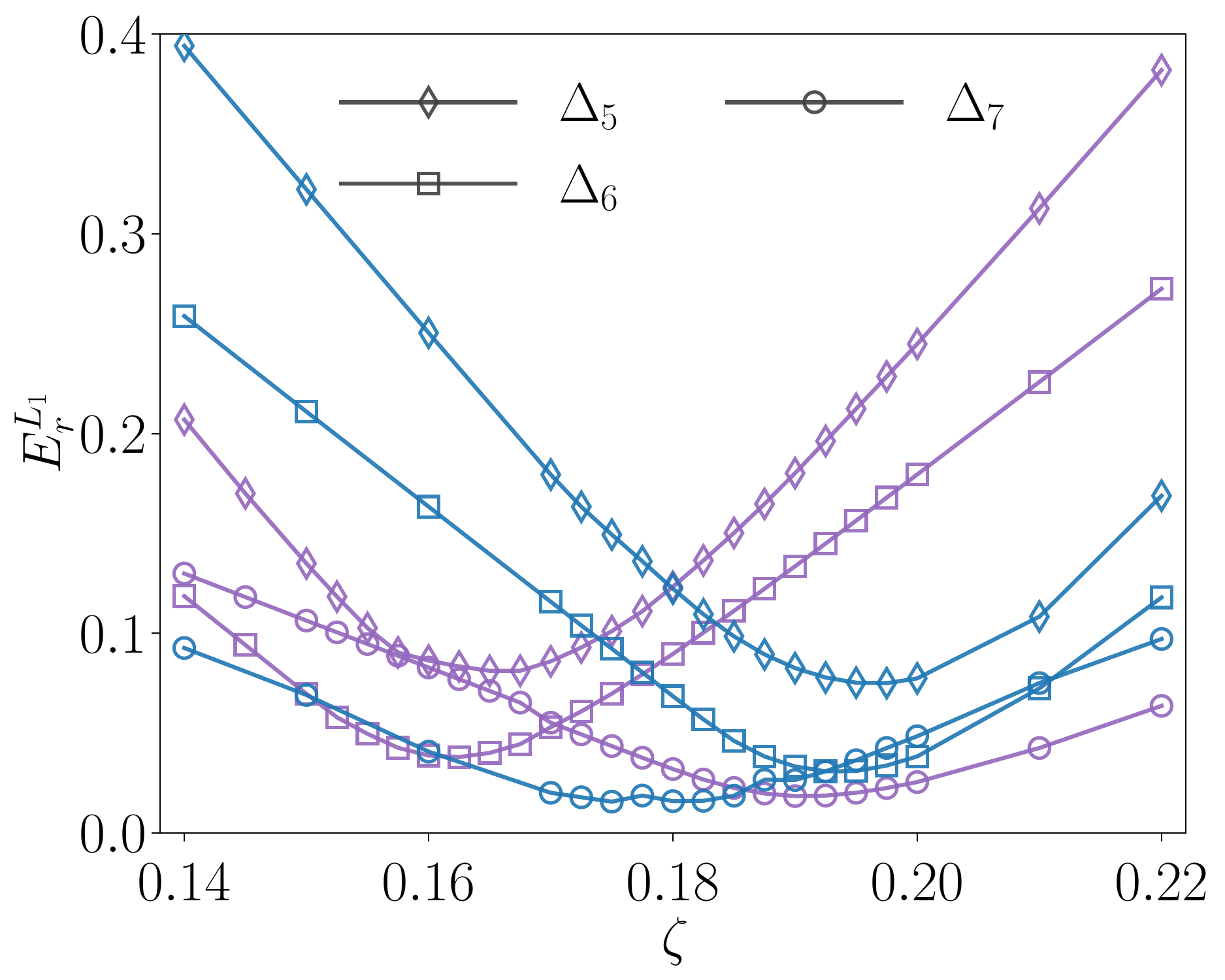}
    \end{minipage}
    \hspace{8mm}
    \begin{minipage}[t]{.46875\columnwidth}
      \centering
      {\small (b)}\\[1mm]
      \includegraphics[width=\linewidth]{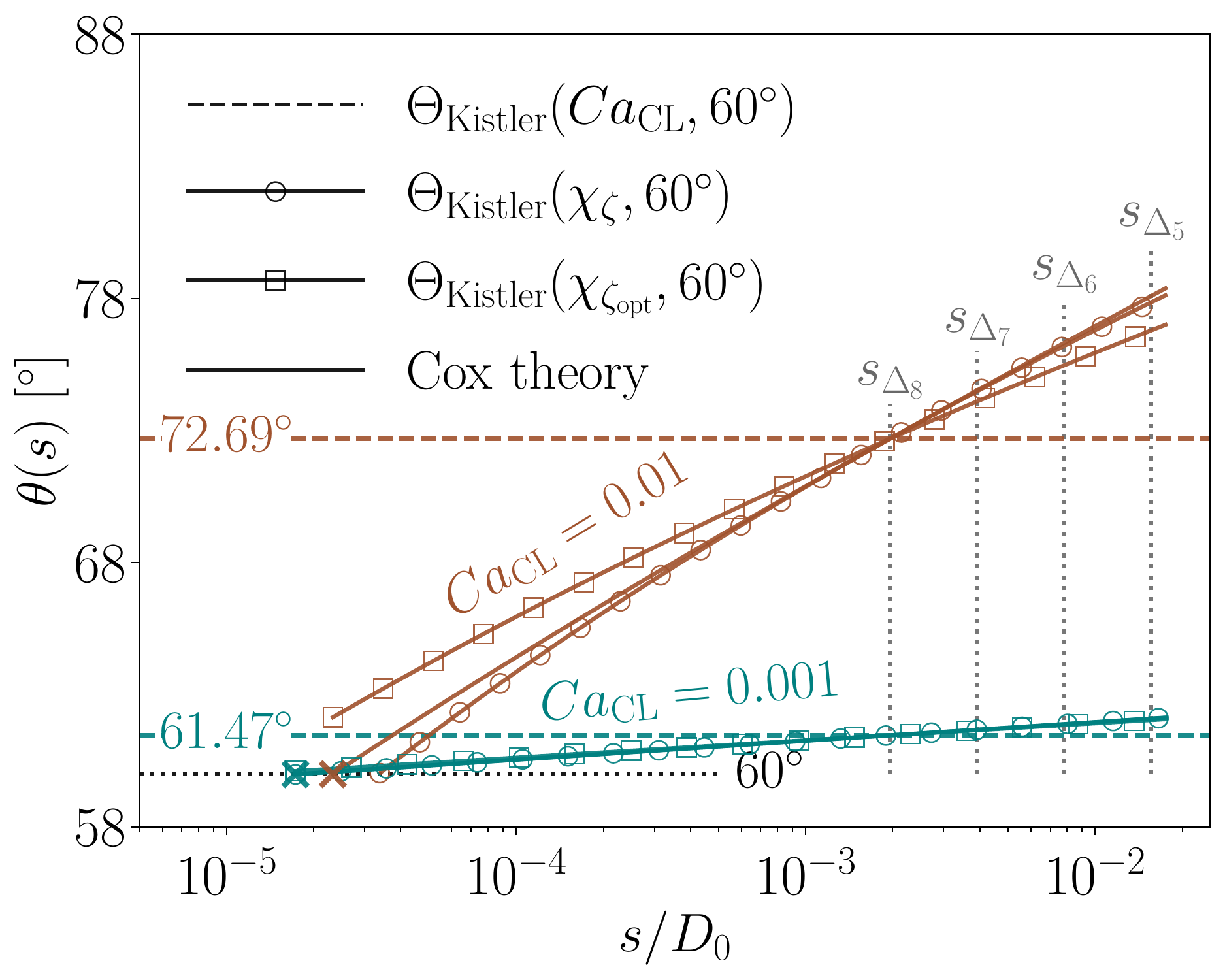}
    \end{minipage}
  }

  \caption{Plots for case `B' (Sec. \ref{sec:caseB}, $\theta_e=60^\circ$) showing (a) variation of the average contact-line position error, $E^{L_1}_r$, with a fixed $\zeta$ for the Kistler model using the contact-line speed calculation proposed by \cite{roisman2008drop} (\textcolor[rgb]{0.502,0,0.502}{\rule[0.5ex]{2.4em}{1pt}}) and the contact-line speed calculation employed in the present work (\textcolor[rgb]{0.1216,0.4667,0.7059}{\rule[0.5ex]{2.4em}{1pt}}), and (b) comparison between the contact angles obtained from the Cox theory (Eq.~\eqref{eq:COX_final}) and from the original and presently `modified' Kistler dynamic contact-angle model for $Ca_{\mathrm{CL}}=0.01$ (\textcolor[rgb]{0.6275,0.3216,0.1765}{\rule[0.5ex]{2.4em}{1pt}}) and $Ca_{\mathrm{CL}}=0.001$ (\textcolor[rgb]{0,0.5,0.5}{\rule[0.5ex]{2.4em}{1pt}}). The distances $s_{\Delta_i}$ correspond to the locations of the cell centers of different grids with sizes $\Delta_i$.
}
  \label{fig:afkh_compare_zeta}
\end{figure}

Fig.~\ref{fig:afkh_compare_zeta}(a) shows that the average error in the contact line position (${E_r}^{L_1}$) attains a local minimum for a certain value of $\zeta (\equiv \zeta_{opt})$ for all the grid sizes. Moreover, this minimum is attained in the neighborhood of $\zeta_{opt} \approx 0.193$ for different grid-sizes, with an average error of ${E^{L_1}_r}\approx 0.047\%$ across the three grid levels. However, the figure also shows that we may not obtain the same $\zeta_{opt}$ if the contact line velocity is approximated by some other method, such as that obtained from an interpolation of cell-centered fluid velocities due to \citet{roisman2008drop}, for which $\zeta_{opt}\approx 0.167$ with an average error of ${E^{L_1}_r}\approx 0.064\%$ across the three grid levels is obtained. The modified Kistler model shows better agreement with Cox's theory for the grid-scale contact angle in contrast to the scale independent, original Kistler model, as shown in Fig.~\ref{fig:afkh_compare_zeta}(b). Now, we proceed with the grid-independence study for the problems of a withdrawing plate and a spreading droplet. 

\subsection{Withdrawing plate case with the present modification}
\label{subsub:Withdrawing plate problem with modified models}

The withdrawing plate problem, where a steadily moving, partially immersed, vertical, flat plate causes a receding contact-line motion, presents a case where $Ca_{\mathrm{CL}}$ attains a constant value corresponding to the plate velocity when $Ca_{\mathrm{CL}}$ is below a critical value. \citet{afkhami2009mesh} considered a withdrawing plate case with $Re=4$ and $Ca_{\mathrm{CL}}=0.03$, with the length scale as pool length and velocity scale as the plate velocity, and demonstrated grid independent numerical results by prescribing mesh-dependent apparent contact angles at the contact line. The simulations are not repeated here; instead, the numerical contact angle to be imposed is calculated from the presently derived modification of two different models (see Table~\ref{tab:afkhami_angles}). Both the Kistler and Tanner models are considered, with $\Delta_0=1/512$ as the base resolution. In absence of the experimental results, for each model and case, $\theta_r$ is determined from the apparent angle at $\Delta_0$, while $\zeta$ is determined from that at $\Delta=1/32$. Since $Ca_{\mathrm{CL}}$ is fixed and $\zeta$ is independent of grid resolution, the same value of $\zeta$ is used at all intermediate resolutions. This gives $(\theta_r,\zeta)=(107.08^\circ,0.335)$ and $(123.87^\circ,0.554)$ for the Kistler model, and $(104.32^\circ,0.425)$ and $(123.59^\circ,0.668)$ for the Tanner model, for the $90^\circ$ and $114^\circ$ base cases, respectively.

\begin{table}
  \begin{center}
\def~{\hphantom{0}}
\small
\renewcommand{\arraystretch}{1.20}
\setlength{\tabcolsep}{5pt}

\begin{tabular}{
@{}c@{}
  @{}p{14pt}@{}
  c
  >{\centering\arraybackslash}p{48pt}
  >{\centering\arraybackslash}p{48pt}
  @{}p{18pt}@{}
  c
  >{\centering\arraybackslash}p{48pt}
  >{\centering\arraybackslash}p{48pt}
  @{}
}

&
&
\multicolumn{3}{c}{$\theta_d(\Delta_0)=90^\circ$}
&
&
\multicolumn{3}{c}{$\theta_d(\Delta_0)=114^\circ$}
\\[5pt]

$\Delta$
&
&
Afkhami \textit{et al.} &
`modified' Kistler &
`modified' Tanner
&
&
Afkhami \textit{et al.} &
`modified' Kistler &
`modified' Tanner
\\[6pt]

$1/512$
&
&
$90^\circ$ & $90.00^\circ$ & $90.00^\circ$
&
&
$114^\circ$ & $114.00^\circ$ & $114.00^\circ$
\\

$1/256$
&
&
$84^\circ$ & $84.69^\circ$ & $84.78^\circ$
&
&
$108^\circ$ & $109.15^\circ$ & $108.95^\circ$
\\

$1/128$
&
&
$78^\circ$ & $78.67^\circ$ & $78.81^\circ$
&
&
$102^\circ$ & $103.72^\circ$ & $103.39^\circ$
\\

$1/64$
&
&
$72^\circ$ & $71.61^\circ$ & $71.78^\circ$
&
&
$97^\circ$ & $97.41^\circ$ & $97.16^\circ$
\\

$1/32$
&
&
$63^\circ$ & $63.00^\circ$ & $63.00^\circ$
&
&
$90^\circ$ & $90.00^\circ$ & $90.00^\circ$
\\

\end{tabular}

\caption{Apparent contact angles reported by~\citet{afkhami2009mesh} for the withdrawing-plate problem
and those obtained using the presently `modified' subgrid models based on the Kistler and Tanner models at
different grid resolutions.}
\label{tab:afkhami_angles}

  \end{center}
\end{table}

Both models closely recover the angles reported by \citet{afkhami2009mesh} at intermediate resolutions. The maximum deviations are less than $1^\circ$ for the $90^\circ$ case and approximately $1.7^\circ$ and $1.4^\circ$ for Kistler and Tanner, respectively, for the $114^\circ$ case.

Thus, a single value of $\zeta$ recovers the mesh-dependent apparent angles across all resolutions for each case, with the scale dependence of the contact angle entering through $\chi$. Since these apparent angles were shown by \citet{afkhami2009mesh} to yield grid-independent solutions, their recovery by the presently `modified' contact-angle models confirm the suitability of the present method for obtaining grid-independent results of macroscopic flows with moving contact lines.

\subsection{Spreading droplet simulations with the present modification}
\label{subsub:Case B: Liquid-liquid system with modified models}
\begin{figure}
  \centerline{%
    \begin{minipage}[t]{.46875\columnwidth}
      \centering
      {\small (a)}\\
      \includegraphics[width=\linewidth]{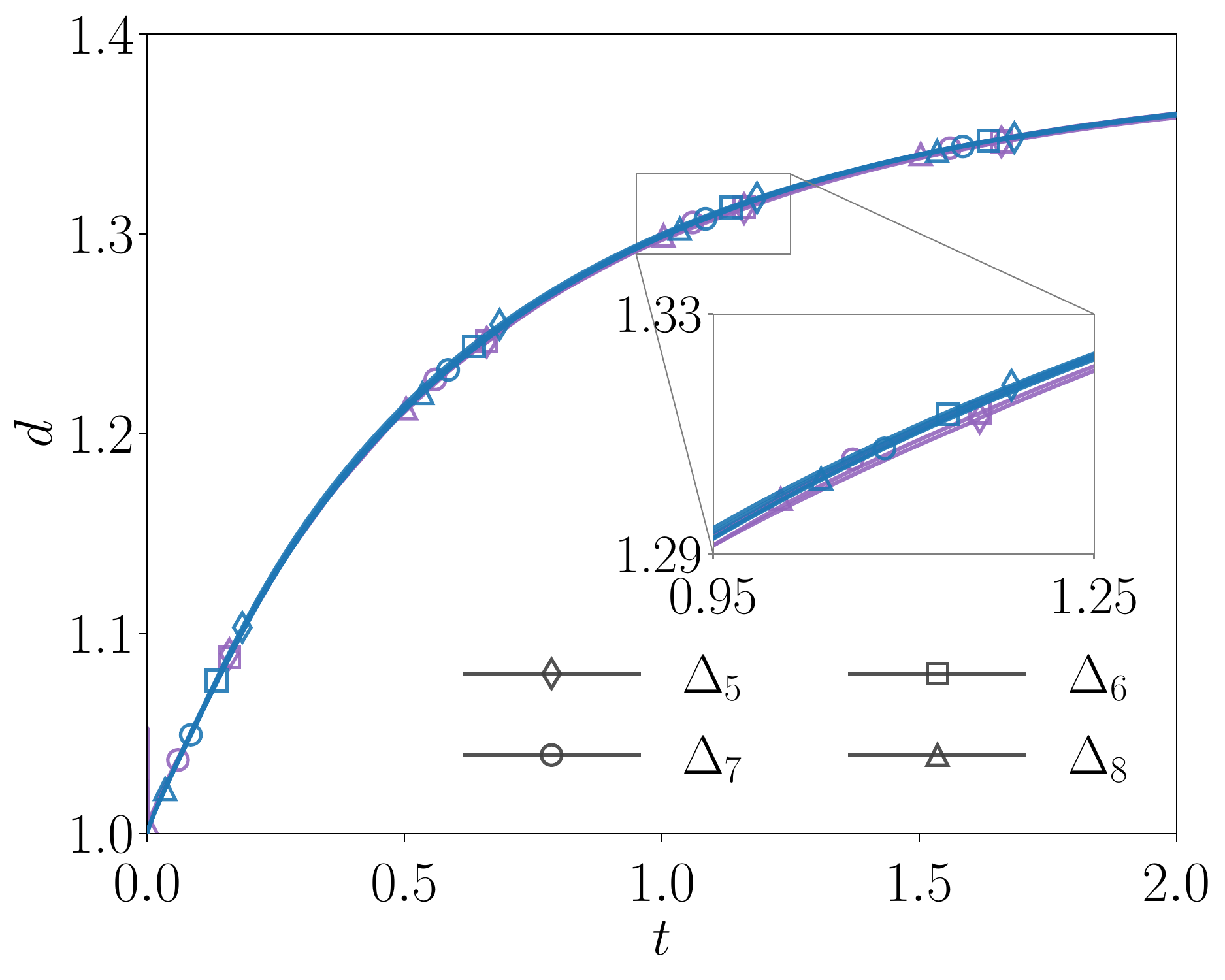}
    \end{minipage}
    \hspace{8mm}
    \begin{minipage}[t]{.46875\columnwidth}
      \centering
      {\small (b)}\\
      \includegraphics[width=\linewidth]{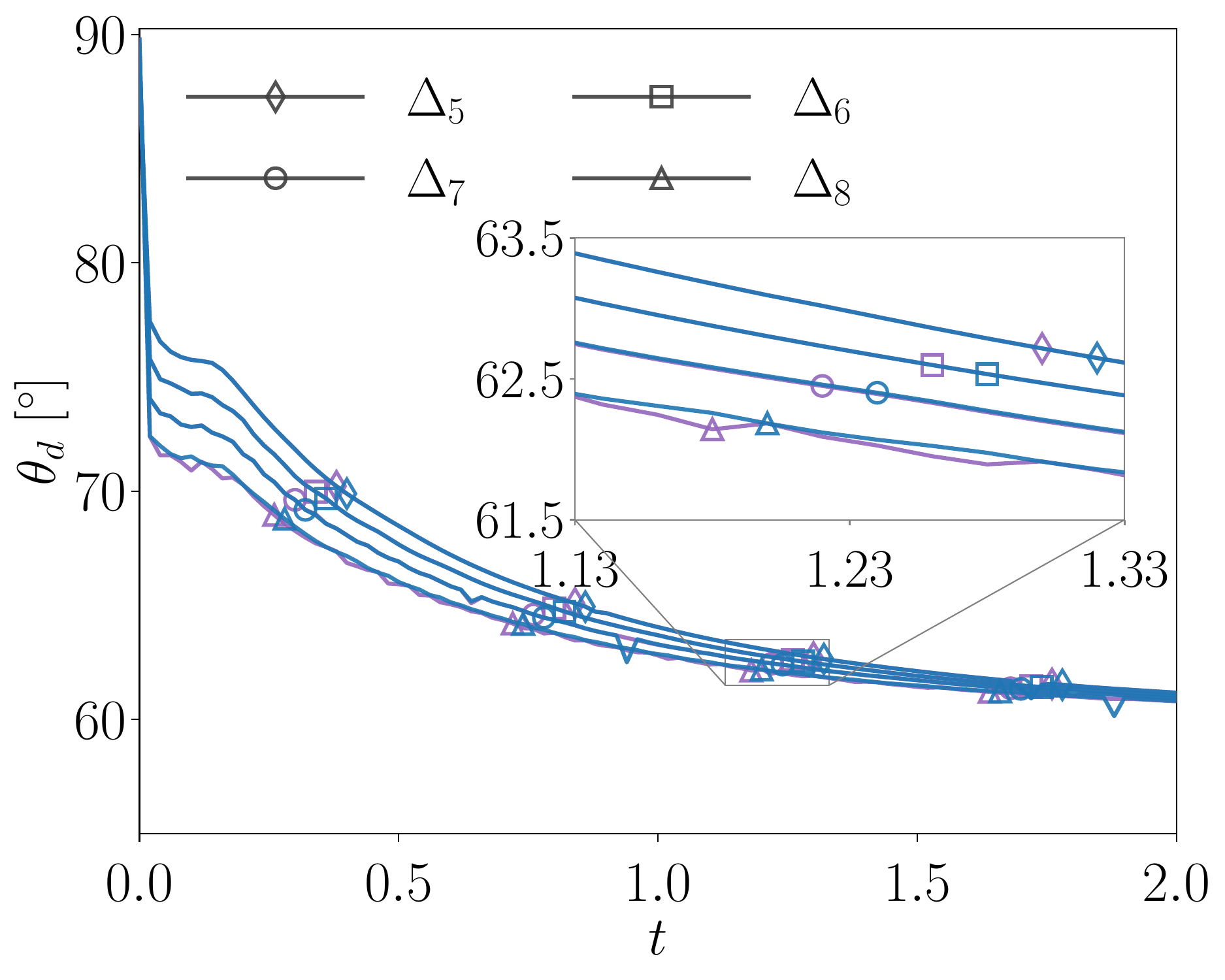}
    \end{minipage}
  }

  \vspace{2mm}
  \centerline{%
    \begin{minipage}[t]{.46875\columnwidth}
      \centering
      {\small (c)}\\
      \includegraphics[width=\linewidth]{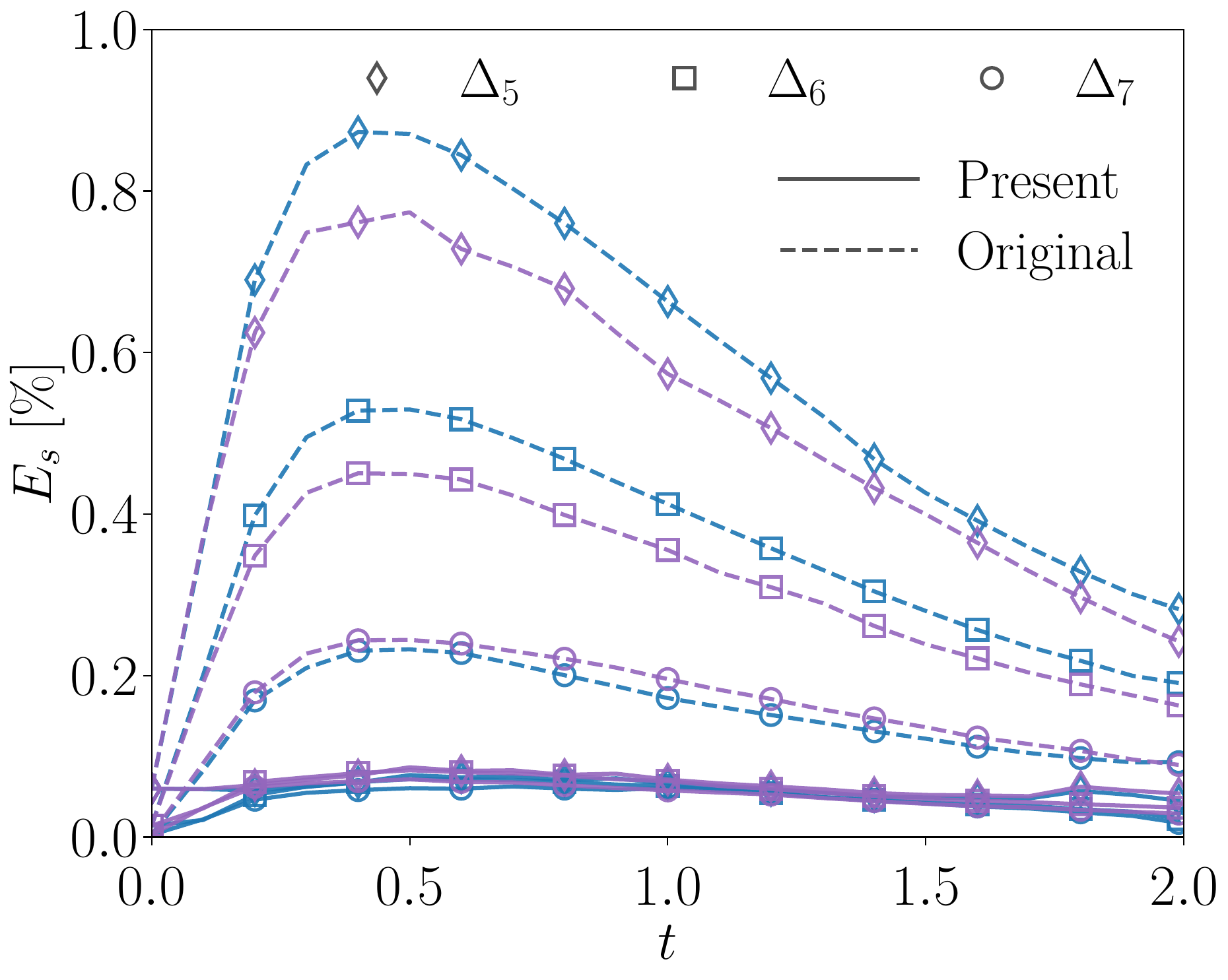}
    \end{minipage}
    \hspace{8mm}
    \begin{minipage}[t]{.46875\columnwidth}
      \centering
      {\small (d)}\\
      \includegraphics[width=\linewidth]{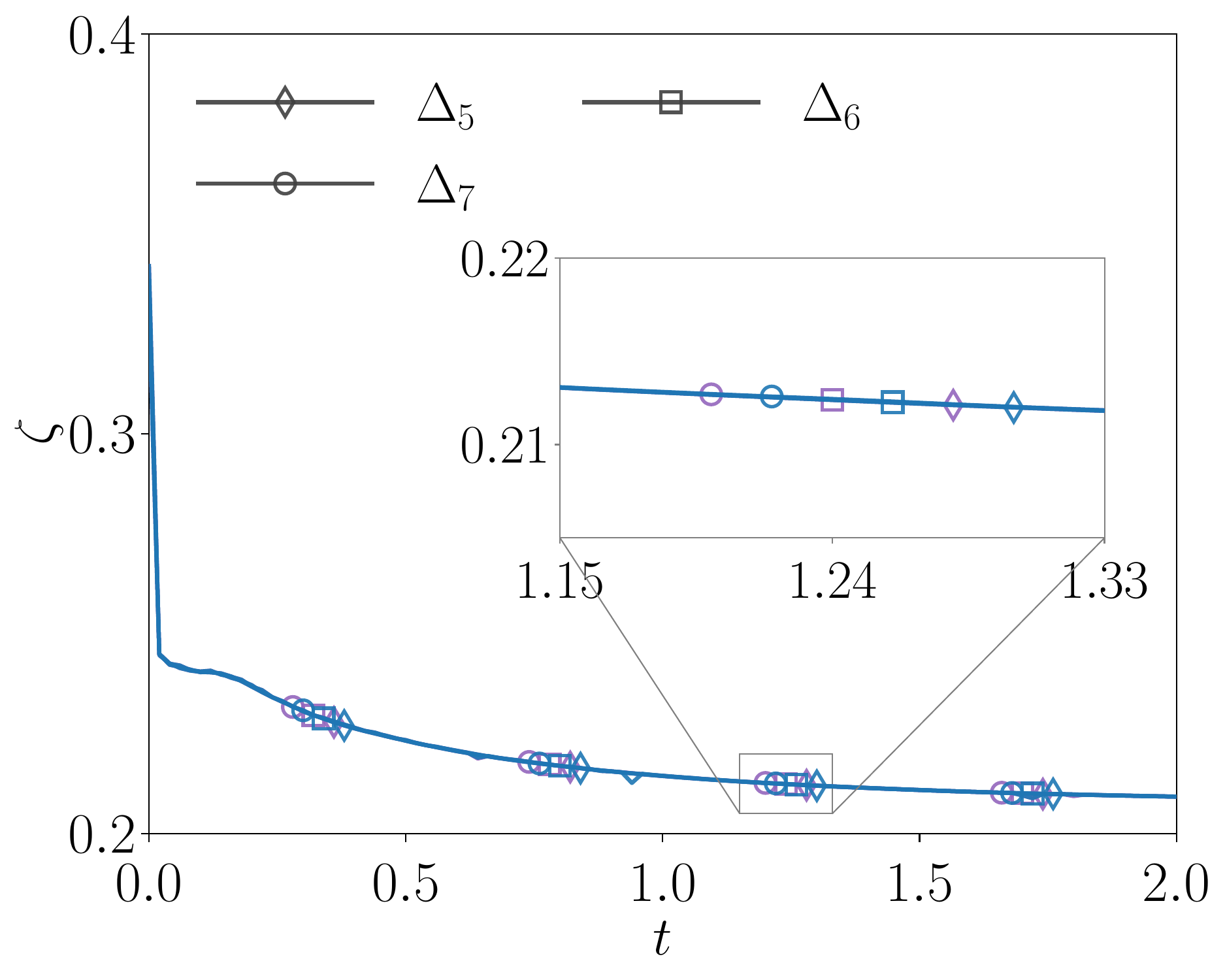}
    \end{minipage}
  }

  \caption{Plots obtained from the present simulations of the liquid-liquid system, case `B' (Sec. \ref{sec:caseB}) with the proposed subgrid model, showing the temporal evolution of (a) spreading diameter ($d$), (b) dynamic contact angle ($\theta_d$), (c) droplet-shape error evaluated relative to the reference grid $\Delta_8$, $E_s^{L_1}$ for the present subgrid model and the original Kistler model, and (d) $\zeta$, for different grid-sizes. The presently `modified' Kistler model has been employed as the subgrid model in all the simulations, with the contact-line speed model proposed in the present work, and the contact-line speed model proposed by \cite{roisman2008drop} represented by \textcolor[rgb]{0.1216,0.4667,0.7059}{\rule[0.5ex]{2.4em}{1pt}} and \textcolor[rgb]{0.502,0,0.502}{\rule[0.5ex]{2.4em}{1pt}}, respectively.
}
  \label{fig:zeta_rcl}
\end{figure}

We run similar numerical experiments at different grid-sizes for case `B' with a dynamically calculated $\zeta$ from Eq.~\ref{eq:zeta_def}. To obtain $\zeta$, we need the observation length scale, $s_0$, of the chosen contact angle model. Once validated against experimental results, the dimensional value of $s_0$ is fixed for a contact angle model. In the absence of a rigorously confirmed value of $s_0$, we assume the observation length scale, $s_0$, of the original empirical contact angle models (Kistler, Jiang, and Tanner) to be approximately $2\mu m$. For a droplet with $D_0 = 4mm$, the corresponding value of $s_0/D_0$ approximately matches the grid-size $\Delta_{8}$. Fig. \ref{fig:zeta_rcl}(a) demonstrates grid-independence of the contact line position obtained from simulations with the presently `modified' contact angle models at different grid sizes. As expected, the imposed contact angles at different grid resolutions must depend on the grid-size due to the viscous bending of the interface (Fig. \ref{fig:zeta_rcl}(b)), while $\zeta$ remains approximately equal to $0.21$ for most of the droplet dynamics (Fig. \ref{fig:zeta_rcl}(c)). Notably, the error in the interface shape is less than $\approx 0.1\%$ at all times, as shown in Fig. \ref{fig:zeta_rcl}(d), demonstrating the grid-independence achieved by using the presently `modified' contact angle model. 

Although a reduction of grid-dependent errors for a similar problem is demonstrated in the literature \citep{afkhami2009mesh,legendre2015comparison}, the authors achieve this by ad-hoc assumptions for the microscopic parameters, $\theta_w$ and $s_m$, and fitting parameters are required to be obtained from the experimental results. In contrast, the present method eliminates the need for fixing $\theta_w$ and $s_m$, and is framed for a general problem to which the Cox theory and the chosen contact angle model are applicable.

We noted in Sec. \ref{subsec:gasliquid} that our numerical method does not require any modification for gas-liquid systems due to negligible errors caused due to grid-sensitivity of the contact line motion. Therefore, next, we focus on the applicability of the modified contact angle model in achieving grid-independence for $\rho_r=1$ and $\mu_r=1$ for a range of $Re$, $We$, and contact angles.

\section{Results and discussion}
\label{Results_discussion}
\subsection{Applicability across $Re$ and $We$ regimes and contact angle models}
In this section, we examine the applicability of the preceding analysis for a range of $Re$ and $We$ by conducting axisymmetric simulations of a droplet spreading on a flat solid surface, with an initial dimensionless velocity $1$ in the downward direction, for two different $\theta_e \in \{60^\circ, 120^\circ\}$, with $\rho_r = 1$, and $\mu_r = 1$. We consider the initial shape of the droplet to be a hemisphere. The droplet recedes or advances, in general, depending on $\theta_e$. Apart from fixing the observation length scale to $\Delta_9$, which can be derived once for an empirical contact angle model, no other tuning parameters are introduced in the present subgrid model.

\begin{figure}
\centerline{%
\begin{minipage}[c]{0.52\columnwidth}
\centering
{\small (a)}\\[1mm]
\includegraphics[width=\linewidth]
{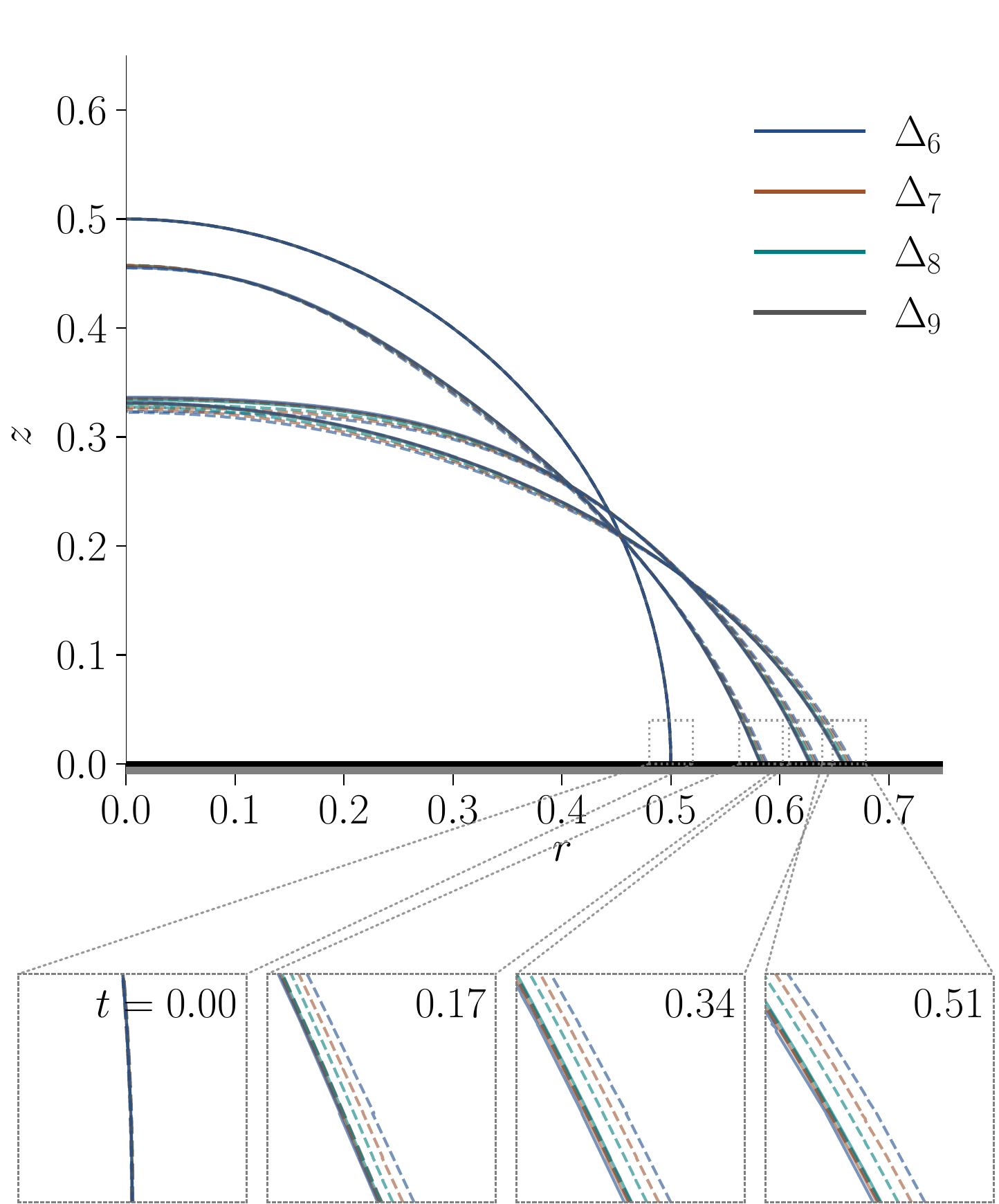}
\end{minipage}
\hspace{8mm}
\begin{minipage}[c]{0.35\columnwidth}
\centering

  {\small (b)}\\[1mm]
  \includegraphics[width=\linewidth]
  {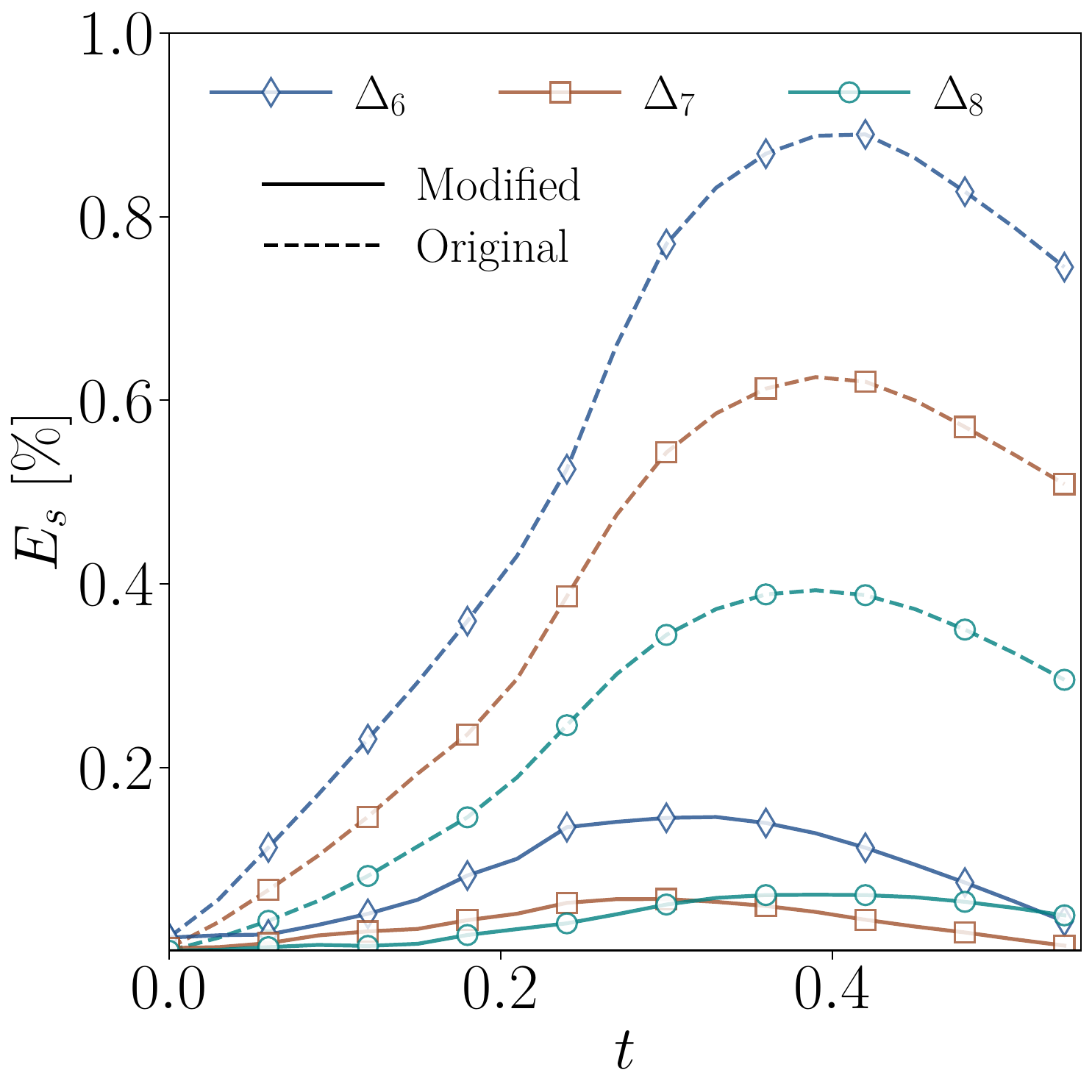}

  {\small (c)}\\[1mm]
  \includegraphics[width=\linewidth]
  {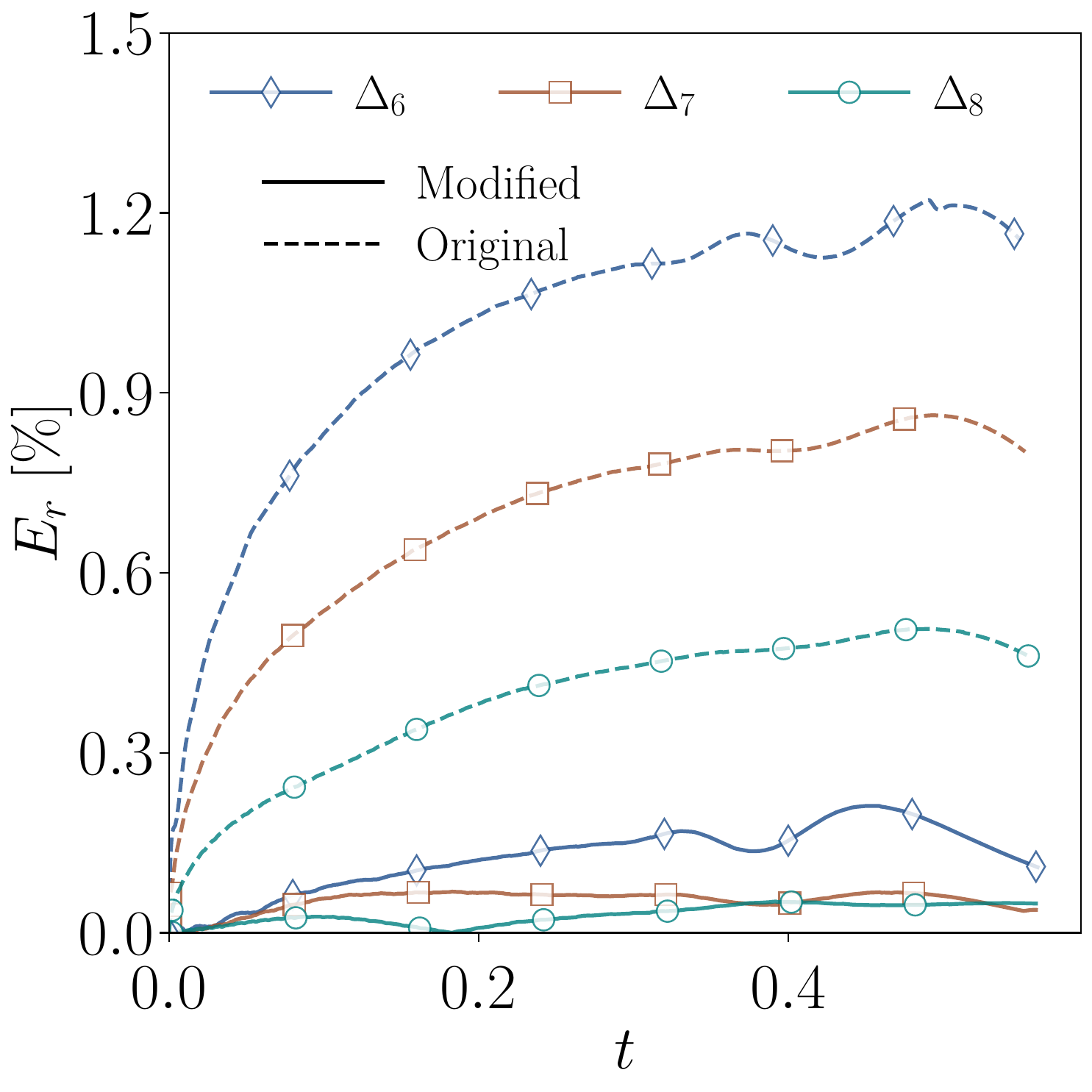}

\end{minipage}

}

\caption{Comparison of (a) droplet interface shapes, (b) droplet-shape error $E$, and (c) contact-line position error $E^{L_1}_r$ between the original and the presently `modified' Kistler dynamic contact-angle model at different grid resolutions ($\Delta_6,\Delta_7,\Delta_8,\Delta_9$) for the base case ($Re=100$, $We=1$, $\rho_r=1$, $\mu_r=1$), and $\theta_e=60^\circ$. The errors in panels (b) and (c) are evaluated relative to the reference grid-size $\Delta_9$.}
\label{fig:errorVsZeta_60}
\end{figure}

\begin{figure}
  \centerline{%
    \begin{minipage}[c]{0.52\columnwidth}
      \centering
      {\small (a)}\\[1mm]
      \includegraphics[width=\linewidth]
      {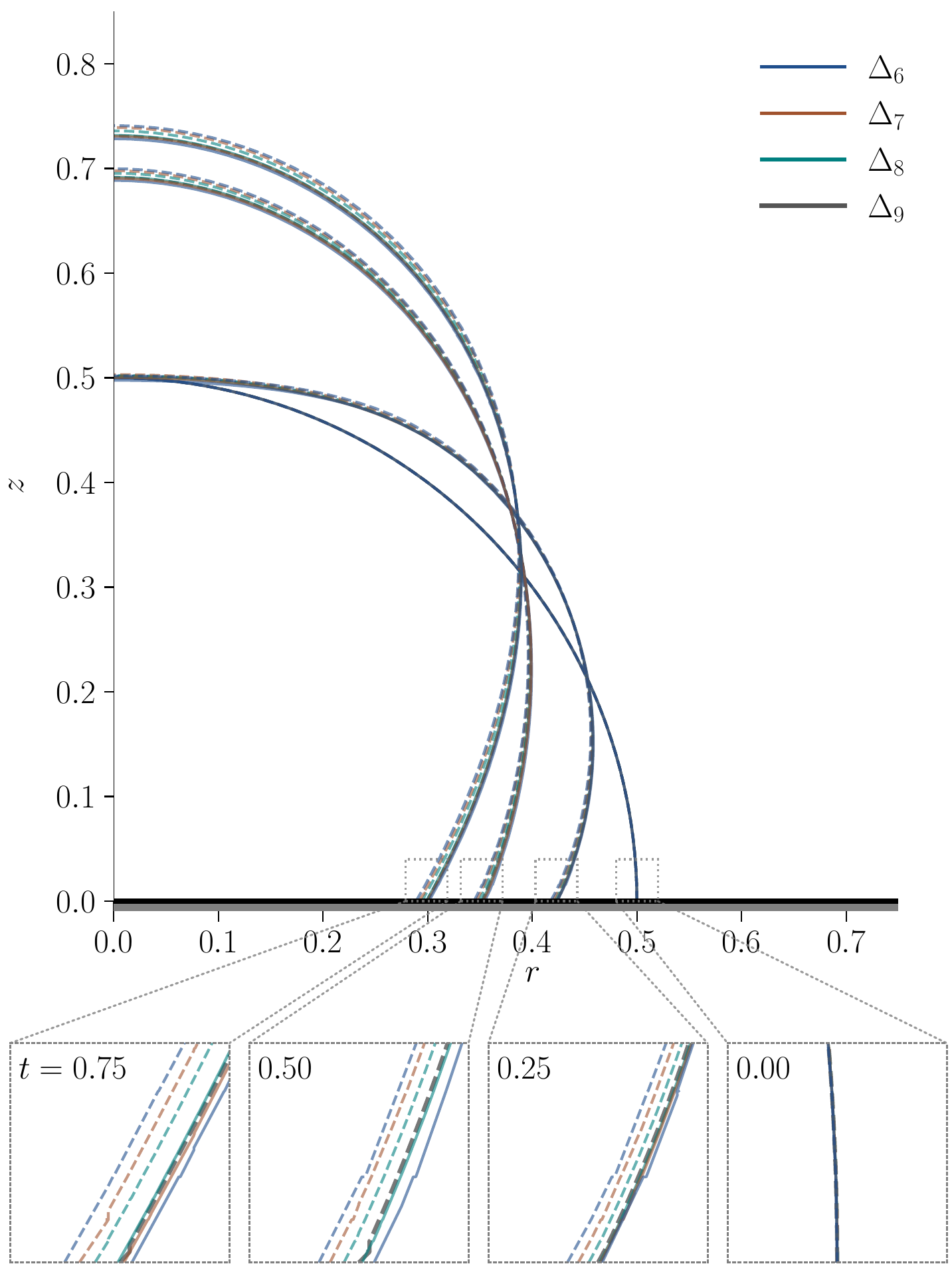}
    \end{minipage}
    \hspace{8mm}
    \begin{minipage}[c]{0.35\columnwidth}
      \centering

      {\small (b)}\\[1mm]
      \includegraphics[width=\linewidth]
      {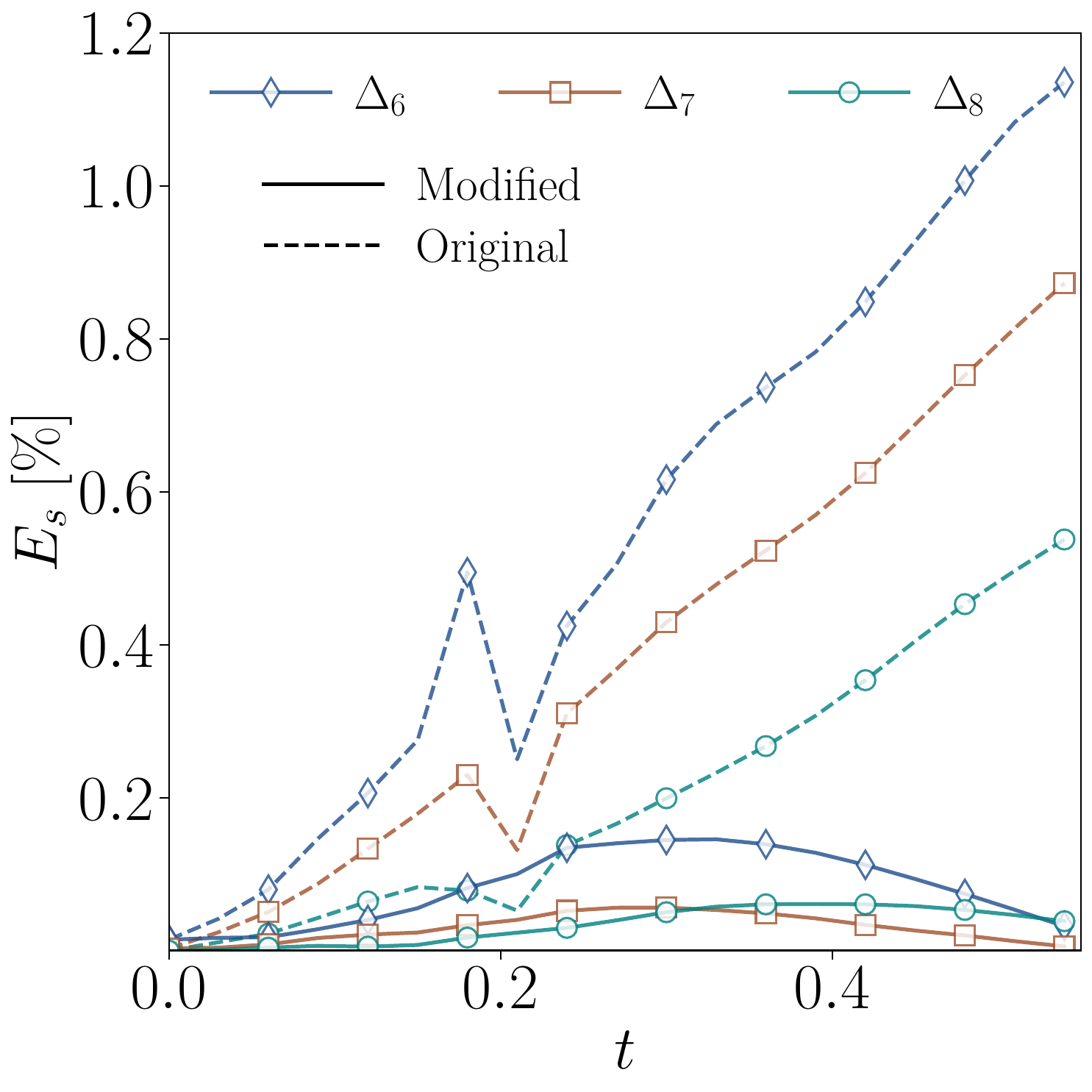}

      {\small (c)}\\[1mm]
      \includegraphics[width=\linewidth]
      {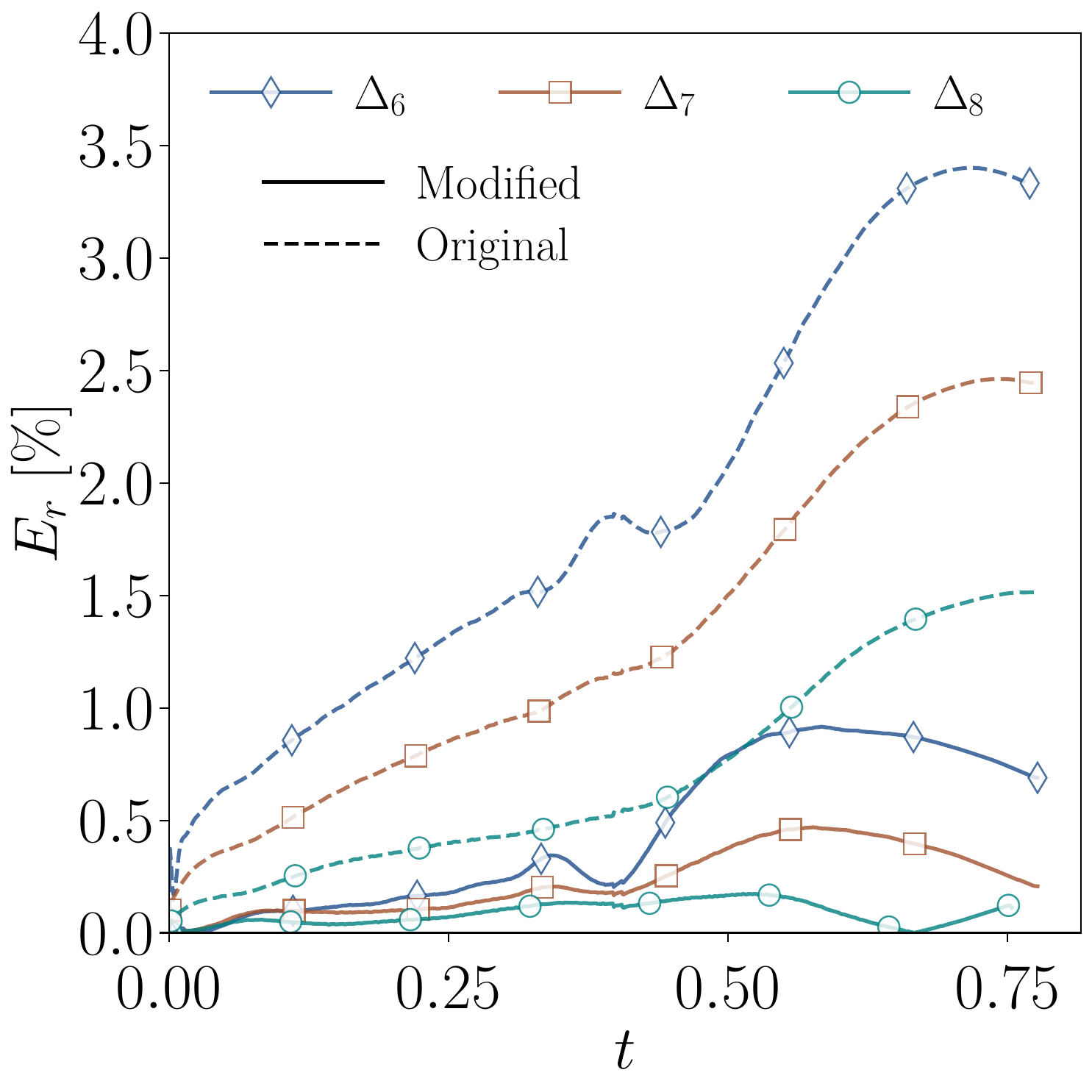}

    \end{minipage}
  }

  \caption{Comparison of (a) droplet interface shapes, (b) droplet-shape error $E$, and (c) contact-line position error $E^{L_1}_r$ between the original and the presently `modified' Kistler dynamic contact-angle model at different grid resolutions ($\Delta_6,\Delta_7,\Delta_8,\Delta_9$) for the base case ($Re=100$, $We=1$, $\rho_r=1$, $\mu_r=1$), and $\theta_e=120^\circ$. The errors in panels (b) and (c) are evaluated relative to the reference grid-size $\Delta_9$.
}
  \label{fig:errorVsZeta_120}
\end{figure}

For $Re=100$ and $We=1$, termed the `base' parameters hereafter, Figs. \ref{fig:errorVsZeta_60} and \ref{fig:errorVsZeta_120} show advancing and receding contact line dynamics, respectively. A temporal deviation in the interface shape, with respect to $\Delta_9$ grid, is observed for all the grids with the original Kistler's model. In contrast, the grid-independence, not just of the contact-line position, but the entire droplet interface shape (Figs. \ref{fig:errorVsZeta_60}(a) and \ref{fig:errorVsZeta_120}(a)) is achieved by employing the presently `modified' Kistler's model. The temporal variation of $E_r$ and $E_s$ suggest about an order of magnitude reduction in the errors (Figs. \ref{fig:errorVsZeta_60}(b) and (c), and \ref{fig:errorVsZeta_120}(b) and (c)). A rapid increase in $E_r$ during the initial times for the `original' model and a corresponding reduction in the error by the presently `modified' model is clearly visible in Figs. \ref{fig:errorVsZeta_60}(c) and \ref{fig:errorVsZeta_120}(c), for the advancing and receding conditions.

The mean relative percentage errors in the contact line position and interface shape between the grid sizes $\Delta_9$ and $\Delta_7$ with the original and the presently `modified' contact angle models are shown in Tables \ref{tab:kistler_jiang_theta60_We1} and \ref{tab:kistler_tanner_theta120_We1} for $\theta_e = 60^\circ$ and $120^\circ$, respectively, for $Re$ ranging from $1$ to $200$. These errors are reduced by approximately an order of magnitude for $Re>=10$ with the modified Kistler and Jiang models; Jiang model performing consistently better across the range of Reynolds numbers considered. A larger error at low $Re$ suggests that higher order approximations in $Ca_{\mathrm{CL}}$ (refer Sec.~\ref{sec:coupling_cox}) are required for this regime where the capillary number is relatively high. We employ Kistler and Tanner contact angle models for $\theta_w = 120^\circ$ as the Jiang's model is applicable only for the advancing contact lines. A similar pattern of error reduction with the presently `modified' models is observed for the entire range of $Re$ in these obtuse angle cases. The errors are generally larger for $\theta_e=120^\circ$ in comparison to the acute angle cases, with and without the proposed modification in the contact angle models.

\begin{table}
  \begin{center}
\def~{\hphantom{0}}
\small
\renewcommand{\arraystretch}{1.20}
\setlength{\tabcolsep}{5pt}

\begin{tabular}{
@{}c@{}
@{}p{18pt}@{}
cc
@{}p{14pt}@{}
cc
@{}p{24pt}@{}
cc
@{}p{14pt}@{}
cc
@{}
}

&
&
\multicolumn{5}{c}{Kistler model}
&
&
\multicolumn{5}{c}{Jiang model}
\\[5pt]

&
&
\multicolumn{2}{c}{Original}
&
&
\multicolumn{2}{c}{Modified}
&
&
\multicolumn{2}{c}{Original}
&
&
\multicolumn{2}{c}{Modified}
\\[5pt]

$Re$
&
&
$E^{L_1}_s$
&
$E^{L_1}_r$
&
&
$E^{L_1}_s$
&
$E^{L_1}_r$
&
&
$E^{L_1}_s$
&
$E^{L_1}_r$
&
&
$E^{L_1}_s$
&
$E^{L_1}_r$
\\[6pt]

1   & & 0.347 & 0.624 & & 0.212 & 0.399 & & 0.296 & 0.522 & & 0.160 & 0.300 \\
4   & & 0.387 & 0.697 & & 0.114 & 0.217 & & 0.230 & 0.407 & & 0.062 & 0.120 \\
10  & & 0.401 & 0.723 & & 0.065 & 0.125 & & 0.205 & 0.364 & & 0.026 & 0.054 \\
40  & & 0.392 & 0.694 & & 0.054 & 0.097 &  & 0.240 & 0.430 &  & 0.012 &  0.028\\
100 & & 0.383 & 0.689 & & 0.031 & 0.054 & & 0.339 & 0.588 & & 0.022 & 0.034 \\
200 & & 0.261 & 0.456 & & 0.065 & 0.073 & & 0.221 & 0.390 & & 0.055 & 0.056 \\

\end{tabular}

\caption{Mean droplet shape errors ($E_s^{L_1}$) and contact-line position errors ($E_r^{L_1}$) for different values of $Re$ for $We=1$ and $\theta_e=60^\circ$ obtained using the original($\zeta=0$) Kistler and Jiang dynamic contact-angle models, and the present subgrid models based on the `modified' Kistler and Jiang models.}
\label{tab:kistler_jiang_theta60_We1}

  \end{center}
\end{table}

\begin{table}
  \begin{center}
\def~{\hphantom{0}}
\small
\renewcommand{\arraystretch}{1.20}
\setlength{\tabcolsep}{5pt}

\begin{tabular}{
@{}c@{}
@{}p{18pt}@{}
cc
@{}p{14pt}@{}
cc
@{}p{24pt}@{}
cc
@{}p{14pt}@{}
cc
@{}
}

&
&
\multicolumn{5}{c}{Kistler model}
&
&
\multicolumn{5}{c}{Tanner model}
\\[5pt]

&
&
\multicolumn{2}{c}{Original}
&
&
\multicolumn{2}{c}{Modified}
&
&
\multicolumn{2}{c}{Original}
&
&
\multicolumn{2}{c}{Modified}
\\[5pt]

$Re$
&
&
$E^{L_1}_s$
&
$E^{L_1}_r$
&
&
$E^{L_1}_s$
&
$E^{L_1}_r$
&
&
$E^{L_1}_s$
&
$E^{L_1}_r$
&
&
$E^{L_1}_s$
&
$E^{L_1}_r$
\\[6pt]

1   & & 0.647 & 1.312 & & 0.525 & 1.039 & & 0.566 & 1.159 & & 0.420 & 0.845 \\
4   & & 0.704 & 1.423 & & 0.312 & 0.604 & & 0.815 & 1.648 & & 0.310 & 0.598 \\
10  & & 0.769 & 1.537 & & 0.227 & 0.426 & & 0.867 & 1.960 & & 0.233 & 0.298 \\
40  & & 0.893 & 1.752 & & 0.167 & 0.299 & & 0.916 & 1.793 & & 0.141 & 0.251 \\
100 & & 0.591 & 1.295 & & 0.119 & 0.230 & & 0.515 & 1.167 & & 0.104 & 0.180 \\
200 & & 0.195 & 0.427 & & 0.165 & 0.283 & & 0.190 & 0.392 & & 0.165 & 0.282 \\

\end{tabular}

\caption{Mean droplet shape errors ($E_s^{L_1}$) and contact-line position errors ($E_r^{L_1}$) for different values of $Re$ for $We=1$ and $\theta_e=120^\circ$ obtained using the original($\zeta=0$) Kistler and Tanner dynamic contact-angle models, and the present subgrid models based on the `modified' Kistler and Tanner models.}
\label{tab:kistler_tanner_theta120_We1}

  \end{center}
\end{table}

Next, we examine the mean errors in the contact-line position and interface shape between the results of $\Delta_9$ and $\Delta_7$ grids for different $We$ at $Re=100$. The error in the droplet shape, $E_s^{L_1}$ is reduced by approximately $10$ times for the entire range of $We$, whereas, $E_r^{L_1}$ is reduced by about $5$ times or more for all $We$ except for $10 <We< 100$ for both the contact angle models considered as shown in Tables~\ref{tab:kistler_jiang_theta60_Re100} for $\theta_e=60^\circ$. Although a significant reduction in both $E_s^{L_1}$ and $E_r^{L_1}$ is observed for the obtuse contact angle (see Table \ref{tab:kistler_tanner_Re100_theta120}), the errors are larger for $\theta_w = 120$ as compared to that for the acute contact angle, an observation consistent with that of Table~\ref{tab:kistler_tanner_theta120_We1}. 

Atomistic simulations of water confined between solid surfaces have shown \citep{huang2008water} that the slip length depends on $\theta_e$ as $(1+\cos\theta_e)^{-2}$, indicating enhanced slip on hydrophobic surfaces. This indicates a dependence of $s_m$ on the contact angle. This is one of the reasons why $s_m$ should not be taken as a constant, especially for obtuse angles, even for a single system under consideration, as the contact angle may vary between the receding and advancing angles. This enhanced slip for larger contact angles may be responsible for larger deviations in $\chi$ from $Ca_{\mathrm{CL}}$, requiring the inclusion of higher order terms in the Taylor series expansion of $G(\Theta(\chi))$ in Sec. \ref{sec:coupling_cox} as explored next.

\begin{table}
  \begin{center}
\def~{\hphantom{0}}
\small
\renewcommand{\arraystretch}{1.20}
\setlength{\tabcolsep}{5pt}

\begin{tabular}{
@{}c@{}
@{}p{18pt}@{}
cc
@{}p{14pt}@{}
cc
@{}p{24pt}@{}
cc
@{}p{14pt}@{}
cc
@{}
}

&
&
\multicolumn{5}{c}{Kistler model}
&
&
\multicolumn{5}{c}{Jiang model}
\\[5pt]

&
&
\multicolumn{2}{c}{Original}
&
&
\multicolumn{2}{c}{Modified}
&
&
\multicolumn{2}{c}{Original}
&
&
\multicolumn{2}{c}{Modified}
\\[5pt]

$We$
&
&
$E^{L_1}_s$
&
$E^{L_1}_r$
&
&
$E^{L_1}_s$
&
$E^{L_1}_r$
&
&
$E^{L_1}_s$
&
$E^{L_1}_r$
&
&
$E^{L_1}_s$
&
$E^{L_1}_r$
\\[6pt]

0.5 & & 0.312 & 1.148 & & 0.038 & 0.145 & & 0.271 & 1.154 & & 0.025 & 0.114 \\
1 & & 0.383 & 0.689 & & 0.031 & 0.054 & & 0.339 & 0.588 & & 0.022 & 0.034 \\
2.5 & & 0.487 & 1.524 & & 0.019 & 0.279 & & 0.242 & 1.553 & & 0.008 & 0.251 \\
10  & & 0.462 & 1.752 & & 0.030 & 0.381 & & 0.212 & 1.677 & & 0.012 & 0.309 \\
25  & & 0.450 & 0.780 & & 0.040 & 0.367 & & 0.348 & 1.747 & & 0.026 & 0.429 \\
100 & & 0.760 & 0.751 & & 0.075 & 0.036 & & 0.745 & 0.572 & & 0.099 & 0.123 \\

\end{tabular}

\caption{Mean droplet shape errors ($E_s^{L_1}$) and contact-line position errors ($E_r^{L_1}$) for different values of $We$ for $Re=100$ and $\theta_e=60^\circ$ obtained using the original($\zeta=0$) Kistler and Jiang dynamic contact-angle models, and the present subgrid models based on the `modified' Kistler and Jiang models.}
\label{tab:kistler_jiang_theta60_Re100}

  \end{center}
\end{table}

\begin{table}
  \begin{center}
\def~{\hphantom{0}}
\small
\renewcommand{\arraystretch}{1.20}
\setlength{\tabcolsep}{5pt}

\begin{tabular}{
@{}c@{}
@{}p{18pt}@{}
cc
@{}p{14pt}@{}
cc
@{}p{24pt}@{}
cc
@{}p{14pt}@{}
cc
@{}
}

&
&
\multicolumn{5}{c}{Kistler model}
&
&
\multicolumn{5}{c}{Tanner model}
\\[5pt]

&
&
\multicolumn{2}{c}{Original}
&
&
\multicolumn{2}{c}{Modified}
&
&
\multicolumn{2}{c}{Original}
&
&
\multicolumn{2}{c}{Modified}
\\[5pt]

$We$
&
&
$E^{L_1}_s$
&
$E^{L_1}_r$
&
&
$E^{L_1}_s$
&
$E^{L_1}_r$
&
&
$E^{L_1}_s$
&
$E^{L_1}_r$
&
&
$E^{L_1}_s$
&
$E^{L_1}_r$
\\[6pt]

0.5 & & 0.508 & 1.148 & & 0.093 & 0.145 & & 0.502 & 1.154 & & 0.080 & 0.114 \\
1 & & 0.591 & 1.295 & & 0.119 & 0.230 & & 0.515 & 1.167 & & 0.104 & 0.180 \\
2.5 & & 0.716 & 1.524 & & 0.152 & 0.279 & & 0.726 & 1.553 & & 0.138 & 0.251 \\
10  & & 0.939 & 1.752 & & 0.226 & 0.381 & & 0.838 & 1.677 & & 0.174 & 0.309 \\
25  & & 0.455 & 0.780 & & 0.267 & 0.367 & & 1.102 & 1.747 & & 0.270 & 0.429 \\
100 & & 0.115 & 0.751 & & 0.086 & 0.036 & & 0.105 & 0.572 & & 0.086 & 0.123 \\

\end{tabular}

\caption{Mean droplet shape errors ($E_s^{L_1}$) and contact-line position errors ($E_r^{L_1}$) for different values of $We$ for $Re=100$ and $\theta_e=120^\circ$ obtained using the original($\zeta=0$) Kistler and Tanner dynamic contact-angle models, and the present subgrid models based on the `modified' Kistler and Tanner models.}
\label{tab:kistler_tanner_Re100_theta120}

  \end{center}
\end{table}

\subsection{Second-order correction to the linear approximation}

The formulation leading to Eq.~\eqref{eq:zeta_final} follows from a first-order Taylor expansion of $G(\Theta(\chi))$ about $\chi=Ca_{\mathrm{CL}}$. The accuracy of this approximation depends on the magnitude of the higher-order terms neglected in Eq.~\eqref{eq:taylor_left}. To assess the validity of the linearization, we retain the quadratic term in the same expansion and compare its contribution with that of the linear term. We first denote the coefficient of the linear term of the Taylor series, given in Eq.~\eqref{eq:taylor_second_order}, as $A_1$ and the coefficient of the quadratic term as $A_2$. Mathematically,

\begin{equation}
\left.
\begin{array}{ll}
\displaystyle
A_1
\equiv
\left.
\frac{dG}{d\Theta}
\frac{d\Theta}{d\chi}
\right|_{\chi=Ca_{\mathrm{CL}}},
\\[16pt]
\displaystyle
A_2
\equiv
\frac{1}{2}
\left.
\frac{d}{d\chi}
\left(
\frac{dG}{d\Theta}
\frac{d\Theta}{d\chi}
\right)
\right|_{\chi=Ca_{\mathrm{CL}}}
=
\frac{1}{2}\left[\frac{1}{f(\theta_0,\mu_r)}
\frac{d^2\Theta}{dCa_{\mathrm{CL}}^2}
-
\frac{1}{f^2(\theta_0,\mu_r)}
\frac{\partial f}{\partial\theta}
\left(
\frac{d\Theta}{dCa_{\mathrm{CL}}}
\right)^2\right].
\end{array}
\right\}
\label{eq:A1_A2}
\end{equation}
Equation~\eqref{eq:taylor_second_order} can therefore be written compactly, to the second order in $\delta\chi$, as
\begin{equation}
G(\Theta(\chi(s)))
=
G(\theta_0)+
A_1\delta\chi+A_2
(\delta\chi)^2.
\label{eq:taylor_second_compact}
\end{equation}
Equating \eqref{eq:taylor_second_compact} with \eqref{eq:implicit_final} yields
\begin{equation}
Ca_{\mathrm{CL}}
\ln\left(\frac{s}{s_0}\right)= A_2(\delta\chi)^2+A_1\delta\chi.
\label{eq:quadratic_delta_chi}
\end{equation}
To simplify further, we denote $Ca_{\mathrm{CL}}\ln\left(\frac{s}{s_0}\right)$ as $B$, and the roots of the Equation~\eqref{eq:quadratic_delta_chi} are given by  
\begin{equation}
\delta\chi
=
\frac{-A_1\pm
\sqrt{A_1^2+4A_2B}}
{2A_2}. 
\label{eq:delta_chi_quadratic_root}
\end{equation}
We choose the root that approaches the linear solution ($\delta\chi_{\mathrm{L}}$) in the limit when $A_2\rightarrow0$, which is given by
\begin{equation}
\delta\chi_{\mathrm{L}}
=
\frac{B}{A_1}.
\label{eq:delta_chi_linear_appendix}
\end{equation}
This root can be written in an algebraically equivalent form as
\begin{equation}
\delta\chi
=
\frac{2B}
{A_1+sign(A_2)\sqrt{A_1^2+4A_2B}}.
\label{eq:delta_chi_quadratic_stable}
\end{equation}
Substituting this in Eq.~\eqref{eq:chi(s)} gives
\begin{equation}
\chi(s)
=
Ca_{\mathrm{CL}}
\left[
1+
\zeta_{\mathrm{Q}}
\ln\left(\frac{s}{s_0}\right)
\right],
\label{eq:chi_quadratic}
\end{equation}
where $\zeta_{\mathrm{Q}}$ is given by,
\begin{equation}
\zeta_{\mathrm{Q}}
=
\frac{2}
{A_1+sign(A_2)\sqrt{A_1^2+4A_2B}}.
\label{eq:zeta_quadratic_A}
\end{equation}
Here, it should be noted that, unlike $\zeta(\equiv\zeta_\mathrm{L})$, which is derived only with linear terms of the Taylor series as given in Eq.~\eqref{eq:zeta_def}), $\zeta_\mathrm{Q}$ depends on $s$ and $s_0$, in addition to $Ca_{\mathrm{CL}}$. From Eq.~\eqref{eq:A1_A2}, $\zeta_\mathrm{L}$ can be written as 
\begin{equation}
\zeta_\mathrm{L}=\frac{1}{A_1},
\label{eq:zeta_A1}
\end{equation}
To further investigate the relative importance of the higher order approximation of $\chi(s)$, we may rewrite Equation~\eqref{eq:quadratic_delta_chi} as
\begin{equation}
\epsilon \Upsilon^2 + \Upsilon - 1 = 0.
\label{eq:upsilon_quadratic}
\end{equation}
where $\Upsilon = \zeta_{\mathrm{Q}}/\zeta_{\mathrm{L}}$ and $\epsilon= \frac{A_2B}{{A_1}^2}$. 

For $|\epsilon|\ll 1$, $\Upsilon$ may be approximated as $1-\epsilon$, yielding
\begin{equation}
\frac{\zeta_{\mathrm{Q}}-\zeta_{\mathrm{L}}}
{\zeta_{\mathrm{L}}}
=-\epsilon.
\label{eq:zeta_relative_difference}
\end{equation}

\begin{figure}
  \centerline{%
    \begin{minipage}[t]{.46875\columnwidth}
      \centering
      {\small (b)}\\[1mm]
      \includegraphics[width=\linewidth]{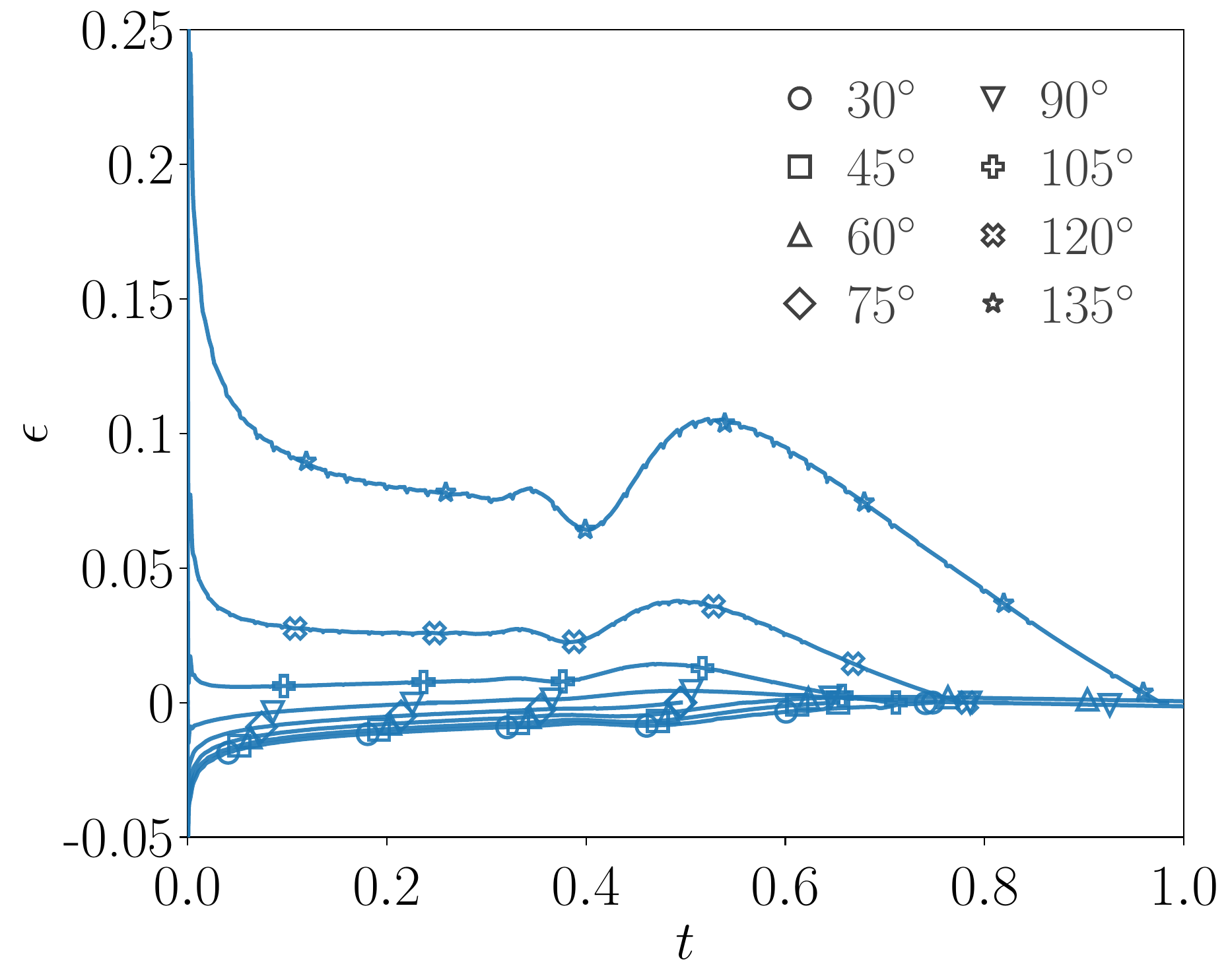}
    \end{minipage}
  }

  \caption{Relative difference ($\epsilon$) between the linear and quadratic approximations of $\zeta$, defined in Eq. (\ref{eq:zeta_relative_difference}), for different equilibrium contact angles ($\theta_e$).
}
  \label{fig:epsilon_rcl}
\end{figure}

Fig. \ref{fig:epsilon_rcl} shows the temporal variation of $\epsilon$ for different contact angles for the `base' parameters. A large deviation from $\epsilon \approx 0$ for contact angles $\ge120^\circ$ points to a need for the higher order correction for obtuse contact angles. 

\begin{table}
\begin{center}
\def~{\hphantom{0}}
\small
\renewcommand{\arraystretch}{1.20}
\setlength{\tabcolsep}{5pt}
\begin{tabular}{
@{}c@{}
@{}p{24pt}@{}
cc
@{}p{24pt}@{}
cc
@{}p{18pt}@{}
cc
@{}
}
& & \multicolumn{2}{c}{Original} & & \multicolumn{5}{c}{Modified} \\[5pt]
& & & & & \multicolumn{2}{c}{Linear} & & \multicolumn{2}{c}{Quadratic} \\[5pt]
$\theta_e$ & & $E^{L_1}_s$ & $E^{L_1}_r$ & & $E^{L_1}_s$ & $E^{L_1}_r$ & & $E^{L_1}_s$ & $E^{L_1}_r$ \\[6pt]
$30^\circ$  & & 0.621 & 1.312 & & 0.050 & 0.096 & & 0.058 & 0.113 \\
$45^\circ$  & & 0.529 & 1.014 & & 0.043 & 0.071 & & 0.049 & 0.085 \\
$60^\circ$  & & 0.383 & 0.690 & & 0.030 & 0.054 & & 0.035 & 0.062 \\
$75^\circ$  & & 0.248 & 0.392 & & 0.016 & 0.032 & & 0.018 & 0.035 \\
$90^\circ$  & & 0.127 & 0.127 & & 0.013 & 0.014 & & 0.013 & 0.014 \\
$105^\circ$ & & 0.210 & 0.394 & & 0.044 & 0.060 & & 0.042 & 0.057 \\
$120^\circ$ & & 0.591 & 1.295 & & 0.119 & 0.230 & & 0.098 & 0.184 \\
$135^\circ$ & & 1.175 & 5.186 & & 0.359 & 1.270 & & 0.235 & 0.778 \\
\end{tabular}
\caption{Mean droplet shape errors ($E_s^{L_1}$) and contact-line position errors ($E_r^{L_1}$) for different values of $\theta_e$ for the `base' parameters ($\rho_r=1$, $\mu_r=1$, $Re=100$, and $We=1$) obtained using the original($\zeta=0$) Kistler model, and the present subgrid model based on the `modified' Kistler model with linear and quadratic approximations of $\zeta$.}
\label{tab:angle_dependence_re100_we1}
\end{center}
\end{table}

Table~\ref{tab:angle_dependence_re100_we1} summarizes the mean contact-line position and interface shape errors for different equilibrium contact angles for the `base' parameters.
Results obtained using the original contact angle models
($\zeta = 0$) show an increase in both $E^{L_1}_r$ and
$E^{L_1}_s$ as the contact angle deviates from $\theta_e = 90^\circ$ and increase rapidly as the
contact angle approaches large obtuse angles.

\section{Conclusion}
\label{sec:conclusion}
In the present work, we propose a subgrid model for numerical simulations of moving contact lines that simultaneously overcomes two bottlenecks, i.e., grid-dependence and the requirement of the knowledge of phenomenological parameters such as the microscopic slip length and wall contact angle. We identify and elaborate on the core reason for grid-dependence as the viscous bending predicted by the Cox theory. Thereafter, We derive a coupling between the Cox theory and the empirical contact angle models that yields the proposed subgrid models based on a modification of the original empirical models, which depend on the distance from the contact line. These `modified' contact angle models are expressed in terms of the observation length scale, i.e., the length scale at which the empirical contact angle model was originally derived. We choose a parameter set from the literature for which grid-dependence is noticeable, and apply our proposed model to achieve grid-independence via a temporal and spatial overlap in the interface shape at different grids. We show the applicability of the proposed method for a range of Reynolds and Weber numbers, and for advancing and receding contact lines. We also show that different methods of estimating the contact line speed yield grid-independent results with the proposed method.

Moreover, it is shown that an apparent match with the experiments may be obtained in some cases even without applying scale-dependent contact angle models as often seen in the literature. This is due to two reasons: an incidental match between the model observation length scale, $s_0$, and the grid-size, and for gas-liquid flows where the variation in the observed contact angle with the distance from the contact-line is comparatively small. In the present implementation of the proposed scale-dependent subgrid model, we do not require a posteriori fine-tuning of the model parameters to match known results, because the only parameter required, $s_0$, needs to be known only once for a contact angle model, which allows predictive simulations of moving contact lines. To our knowledge, this is the first time a framework capable of predictive simulations of such flows, via the numerical implementation of a quasi-universal contact angle model, has been reported.

The proposed modification in the `original' empirical contact angle models is via a multiplicative factor of $Ca_{\mathrm{CL}}$ of the form $1 + \zeta\ln(\Delta/(2s_0))$, for a grid-size $\Delta$. This factor has been assumed to be close to unity for the first-order approximation to be valid. For larger obtuse angles ($\ge120^\circ$), the present first-order approximation of $\zeta$ retains higher errors, which have been shown to reduce when a second-order approximation for $\zeta$ is employed. Moreover, the present model's validity depends on the validity of the Cox theory and that of the `original' empirical contact angle models, without being dependent on $Re$, $We$, $\mu_r$ and $\rho_r$, which encompasses a large spectrum of flows, thus justifying the qualifier `quasi-universal'. However, a more rigorous parametric study needs to be performed to assess the grid-dependence related errors for different parameter regimes.

Additioanlly, although the present simulations are axisymmetric, the scale correction is derived from the local contact-line description and is not specific to axisymmetric geometry. This suggests that the formulation can be extended locally along a three-dimensional contact line, which warrants further examination in the future.

\bibliographystyle{jfm}
\bibliography{references_jfm}

\end{document}